\documentclass{jfm}
\usepackage{graphicx}
\usepackage{epstopdf, epsfig}
\usepackage{placeins}
\usepackage{array}
\usepackage{portland}
\usepackage{times}
\usepackage{enumitem}
\usepackage{epsfig}
\usepackage{subcaption}
\usepackage{amsmath}
\usepackage{float}
\usepackage{xcolor, soul}
\usepackage{physics}
\usepackage{multirow}
\usepackage{pdflscape}
\usepackage[pagewise]{lineno}
\usepackage{natbib}
\usepackage{tabularx}
\newcolumntype{Y}{>{\centering\arraybackslash}X}

\newcommand{\bff}{\mbox{\boldmath $f$}}
\newcommand{\bfv}{\mbox{\boldmath $v$}}
\newcommand{\bfome}{\mbox{\boldmath $\omega$}}

\shorttitle{3-Dimensional Jovian Vortices Constrained by 2-Dimensional Observations}
\shortauthor{A. Zhang and P. S. Marcus}
\title{Stable 3-dimensional Vorticies Consistent with Jovian Observations Including the Great Red Spot}

\author{Aidi Zhang\aff{1},
  Philip S. Marcus\aff{1,2,3}\corresp{\email{pmarcus@me.berkeley.edu}}}

\affiliation{\aff{1}Department of Mechanical Engineering, University of California, Berkeley, CA 94720
\aff{2}Program in Applied Science \& Technology, University of California, Berkeley, CA 94720
\aff{3}Center for Integrative Planetary Science, University of California, Berkeley, CA 94720}

\begin{document}

\maketitle

\begin{abstract}
Detailed observations of the velocities of Jovian vortices exist at only one height in the atmosphere, so their vertical structures are poorly understood. This motivates this study that computes stable 3-dimensional, long-lived planetary vortices that satisfy the equations of motion. We solve the anelastic equations with a high-resolution pseudo-spectral method using the observed Jovian atmospheric temperatures and zonal flow. We examine several families of vortices and find that {\it constant-vorticity} vortices, which have nearly-uniform vorticity as a function of height and horizontal areas that go to zero at their tops and bottoms, converge to stable vortices that look like the Great Red Spot (GRS) and other Jovian anticyclones. In contrast, the {\it constant-area} vortices proposed in previous studies, which have nearly-uniform areas as a function of height and vertical vorticities that go to zero at their tops and bottoms, are far from equilibrium, break apart, and converge to {\it constant-vorticity} vortices. Our late-time vortices show unexpected properties. Vortices that are initially non-hollow become hollow (i.e., have local minima of vertical vorticity at their centers), which is a feature of the GRS that cannot be explained with 2-dimensional simulations. The central axes of the final vortices align with the planetary-spin axis even if they initially align with the local direction of gravity. We present scaling laws for how vortex properties change with the Rossby number and other non-dimensional parameters. We analytically prove that the horizontal mid-plane of a stable vortex must lie at a height above the top of the convective zone.
\end{abstract}

\section{Introduction}
Jupiter and Saturn have robust, long-lived vortices at their mid-latitudes. The Great Red Spot (GRS) of Jupiter, with a current mean diameter of $\sim13,000$~km, has survived for more than 9,600 of its own vortex turn-around times since the time  it  became continuously monitored (starting with \citet{dawes1857observations}), and, if it is the same vortex observed by \citet{hooke1spot} and 
\citet{cassini1more}, then it has survived at least 21,000 turn-around times. The GRS is not unique in its longevity. Three Jovian vortices known as the White Ovals with diameters of $\sim$~9,000~km, each survived for more than 6,000 turn-around times before merging to form the current Red Oval or Little Red Spot. In contrast, the Atlantic Meddies can survive for up to $\sim 150$ vortex turn-around times \citep{richardson2000census}, while terrestrial tropical cyclones over open water typically survive only $\sim3$ vortex turn-around times. More than 200 persistent mid-latitude vortices with diameters greater than 1,000~km have been identified on Jupiter and more than 50 on Saturn \citep{trammell2014global}. These vortices are governed by the familiar laws of fluid dynamics, so their ubiquity and persistence are of interest to and should be explainable by fluid dynamicists, yet there remain many unanswered questions: Why are these vortices so prevalent in some planetary atmospheres but not others? What determines their longevity? Why are observed mid-latitude planetary vortices and Atlantic meddies predominately anti-cyclones, whereas the polar vortices on Jupiter and Saturn, and tropical storms on Earth are cyclonic? For Jovian and Saturnian mid-latitude vortices where the data is available, the vortices are all embedded in east-west zonal flows that have the same sign vorticity as the vortices, but are the zonal flows necessary for their formation or persistence? Must the vortices be continuously forced by underlying thermal convection to persist?

Although we now have high-resolution velocity maps \citep{simon2014dramatic,wong2021evolution} for many of the vortices due to advances in correlation imagery techniques \citep{asay2006extraction,shetty2006modeling,asay2008velocity,asay2009jupiter}, the maps provide only the horizontal (north-south and east-west) velocities and are given at only one height in the atmosphere, the top of the visible clouds (which we have argued is near 1~bar in the atmosphere \citep{marcus2019equatorial}), but others believe to be higher in the atmosphere \citep{fletcher2010thermal,fletcher2016mid,fletcher2021jupiter,wong2021evolution}. We believe that to answer the questions posed above, we need to understand the 3-dimensional structures of these vortices, which to date remain unknown and controversial. For example, it is uncertain whether the GRS lies entirely at heights above the Jovian convective zone or whether its bottom lies deep within it. 
Despite recent advances in observations that are specifically designed to probe multiple depths of the Jovian atmosphere, such as radio measurements with the Very Large Array radio telescope \citep{de2016peering}, gravity field anomaly measurements \citep{parisi2021depth} and microwave \citep{bolton2021microwave} measurements by the {\it Juno} spacecraft, and infrared measurements with the James Webb Space Telescope \citep{de2022jwst}, there is not a consistent 3-dimensional picture of the vortices. We believe that the best way to obtain an understanding of the 3-dimensional vortex dynamics is with numerical simulations such that the simulated vortices are consistent with the horizontal velocities at the one available height where they can be measured in high-resolution and with the 3-dimensional equations of motion.

The few partial answers that have been found to the questions posed above were generally provided by 2-dimensional numerical simulations using reduced equations of motion (shallow-water, quasigeostrophic, etc.) \citep{dowling1988potential,dowling1989jupiter,marcus1993jupiter,showman2007numerical,li2020modeling}. However, these studies beg the question of whether vortices in two dimensions behave as they do in three dimensions (and we show in this paper that in some important and relevant ways that they do not). There have been few numerical calculations of 3-dimensional, Jovian-like vortices and most used the Boussinesq approximation \citep{williams1997planetary,hassanzadeh2012universal,lemasquerier2020remote}, which is appropriate for a fluid without large vertical density variations, but not for the Jovian atmosphere. Even if we use one of the smallest published estimates of the vertical thickness of the GRS, 190~km, the mass density within the GRS varies by a factor of 100 from its top to its bottom.\footnote{\citet{fletcher2010thermal} argue that the top of GRS is at a height above $140$~mbar, while  \citet{parisi2021depth} estimates that the bottom of the GRS is between $150$~km and $375$~km, which corresponds to 40 and~340~bar.  Using 40~bar for the bottom, we obtain that the thickness of GRS is 190~km and then obtain the difference of the densities at the top and bottom from figure~\ref{fig:thermal_bk}.

}
Therefore,  it is not surprising that Boussinesq simulations have velocities that qualitatively differ from the observed GRS velocities at the cloud tops, and the results of these studies may not be applicable to Jovian vortices.   
\citet{liu2010mechanisms} used a global circulation model based on the hydrostatic primitive equations for
a compressible ideal-gas atmosphere, with 30 levels of vertical resolution to compute zonal flows on Jupiter. Their focus was on the east-west jets, not the vortices. During the simulations, vortices formed spontaneously, but they were predominately cyclones, rather than anticyclones. Some numerical 3-dimensional studies used the anelastic equations (or a combination of Boussinesq and anelastic equations) to simulate the Jovian convective zone and the stable-stratified layer at heights above it containing the visible cloud tops, and they were able to create convection-forced 3-dimensional vortices at the cloud-top levels. However, these vortices survived for fewer than 5 turn-around times \citep{heimpel2016simulation,yadav2020deep}.
\citet{cho2001high} solved the 3-dimensional quasigeostrophic equation using contour dynamics and 8 vertical levels. Their focus was on correlating the horizontal contours of their simulation with the cloud patterns of the GRS, and they did not report the vertical structure or the longevity of their vortex. Their calculation was initialized with a large vortex, rather than creating one via convective forcing. The vortex was allowed to freely evolve in time. \citet{morales2013jupiter} followed this line of reasoning to examine a freely evolving vortex to see if the vertical vorticity (as a proxy for the Jovian clouds) would form the observed patterns of the GRS. They solved the primitive isentropic-coordinate
equations (with the EPIC code) initialized with a 3-dimensional ellipsoidal vortex. Their vortex was not forced by convection nor with any diabatic heating from radiative transfer of latent heating or cooling from cloud formation, sublimation, or evaporation.
They successfully produced long-lived vortices. However, they carried out their calculations such that the lower parts of their vortices were in an ambient atmosphere that was much more stably stratified with respect to instability than the Jovian atmosphere, and they did not report on the vertical structures of their vortices (possibly because they only used 14 levels in the vertical direction of their calculation, four of which were sponge layers at the top of the calculation used to prevent numerical instabilities). \citet{palotai20143d} followed up the study of \citet{morales2013jupiter} and modeled how ammonia clouds interact with these vortices. They found ammonia clouds forming inside anticyclones, consistent with Jovian observations. 

In this paper, we follow the approach taken by \cite{morales2013jupiter} and evolve initial vortices, without forcing, diabatic heating, heat transfer, or any form of thermal or viscous dissipation with the equation of motions to see if they survive, qualitatively change, or are destroyed. The goal of this study is to produce non-decaying vortices so that in follow-up studies we can use these vortices to examine their longevity and slow evolution when realistic forcing and dissipation are included. Here, our goal is to examine the vertical structures and other properties of the unforced/undissipated vortices. We use the anelastic equations, which allow for global and local convection (the latter of which we find is important for reproducing several features of Jovian vortices). The calculations are well-resolved in the horizontal and vertical directions, with 385 vertical collocation points. Rather than comparing our simulated vortices to Jovian cloud patterns that they might produce, we compare their horizontal velocities and temperature fields at the cloud top height to observations. The simulated vortices mimic several prominent features of the GRS, including the shielded vorticity, the quiescent interiors with warm, high-velocity annular rings at their peripheries, a cool top, first observed by {\it Voyager} \citep{flasar1981thermal}, and a warm bottom, first proposed by \citet{marcus2013jupiter}.

Our calculations are carried out using the observed atmospheric temperature and pressure \citep{de2019jupiter,moeckel2022ammonia}, which consists of a highly-stratified region at heights above a convection zone (see $\S$ \ref{sec:new} for details). The atmosphere is highly-stratified at the top of our computational domain and has $\bar{N}/f \gg 1$, where $\bar{N}$ is the Brunt-V\"{a}is\"{a}l\"{a} frequency of the atmosphere and $f$ is the local Coriolis frequency. The convection zone at the bottom of our computational domain is near adiabatic ($\bar{N}/f \simeq 0$). 
We include the convection zone in our domain because many researchers expect that the vortex bottom extends nearly to its top, which we set to be 10~bar in our calculations. Some observers believe the bottom of the large Jovian vortices lie within the convection zone \citep{bolton2021microwave}. We want to allow simulations of such scenarios. In addition to the choice of hydrostatic background, our method differs from previous studies because of the numerical method we use (see the Appendix for details). Most numerical methods used for computing atmospheric flows require small time steps when $\bar{N}/f$ is large, making the computation of Jovian vortices numerically challenging. However, our time-stepping algorithm allows us to use a relatively large time step, which makes the simulations feasible with our available computational resources. Using this method we are able to have a large Courant number compared to previous calculations, without sacrificing accuracy or stability. 

Our philosophy for computing stable vortices that look like Jovian vortices incorporates the same set of ideas that we used for computing vortices in non-forced and non-dissipated rotating, stratified Boussinesq flows \citep{hassanzadeh2013unexpected, mahdinia2017stability}. 
Those Boussinesq flows have multiple families of equilibrium vortex solutions, and within each family there is a {\it continuum} of vortex solutions. The fact that each family is a {\it continuum} not only allows the vortices to have different Rossby and Burger numbers, but also allows the vortices to have a continuous range of sizes and a continuous range of locations at different heights within the flow. To compute stable vortices in stratified, rotating Boussinesq flows we used an initial-value code that was initialized with a vortex that was ``close to'' a stable equilibrium, Typically, but not always, if the initial vortex was sufficiently ``close to'' a stable equilibrium, the flows settled into a late-time vortex whose appearance looked similar to the initial condition. 
The approach of the initial condition to equilibrium was facilitated by the fact that the boundaries of the computational domain were designed to absorb, rather than reflect, the Poincar\'e waves that were generated and radiated by the initial vortex as it rapidly adjusted to come into hydrostatic and geostrophic or cyclostrophic equilibrium. 
Because the interior of the flow’s computational domain was dissipationless, once the rapid adjustment and shedding of Poincar\'e  waves ended, the late-time equilibrium vortices stopped dissipating and lasted indefinitely. If the initial condition of the initial-value code was a vortex that was not sufficiently ``close to'' a stable equilibrium, and if the flow remained a vortex at late times, then the late-time vortex did not look like the initial vortex. Frequently, the initial vortex split into multiple vortices, some of which survived, but none of which looked like the initial vortex. Here in this study of Jovian-like vortices in a rotating, stratified Jovian-like atmosphere governed by the anelastic equations of motion, we take the same approach by using an initial-value code and choosing an initial condition that ``looks like’’ a Jovian vortex and is sufficiently ``close to'' a stable equilibrium'', where ``looks like’’ and ``close to'' are defined below.

 One of the goals of this paper is to show how to create an initial 3-dimensional vortex that is sufficiently ``close to'' 
a stable equilibrium of the anelastic equations of motion so that at late times it converges to a vortex similar in appearance to the initial condition and also ``looks like'' a Jovian vortex because its cloud-top velocity is similar in appearance to that of an observed Jovian vortex.%

The remainder of this paper is as follow: In $\S$~2, we review the anelastic equations. In $\S$~3 and~4 we review the Jovian observations we use to determine the initial atmosphere's vertical thermal structure and the structure of the east-west zonal system of winds in which we embed the initial vortex. In $\S$~5 we show how to create initial vortices that are approximate equilibria so that they are sufficiently ``close to '' an equilibrium that they evolve to a stable vortex that looks like the initial condition that itself ``looks like'' a Jovian vortex. In particular, we consider four families of anticyclonic vortices, one of which was a family proposed by \citet{parisi2021depth} as a  model of the GRS to interpret the {\it Juno} gravitometer data. In $\S$~6, we present the results of running the initial-value code with the families of initial anticyclones.  Not all families survive. In this section, we also compare the velocity and temperature fields of the final, quasi-steady vortices to Jovian vortices. In $\S$~7 we determine how various properties of the final vortices scale with their Rossby numbers and with their vertical aspect ratios (the vertical thickness divided by the mean horizontal diameter of the vortex). We also present a semi-analytic formula that explains the shapes of the vortices in their vertical-east/west planes. In $\S$~8 we take a closer look at the vertical structures of the vortices and show how local convective instabilities constrain the locations of the tops and bottoms of the vortices. A summary and discussion of our results, along with a plan future work, are in $\S$~9.

\section{Coordinate Systems, Hydrostatic Equilibrium, and Anelastic Equations of Motion}\label{sec:analytic}

\subsection{Useful Coordinate Systems}\label{sec:analytic_coordinate}
\begin{figure}
    \centering
    \includegraphics[width=\textwidth]{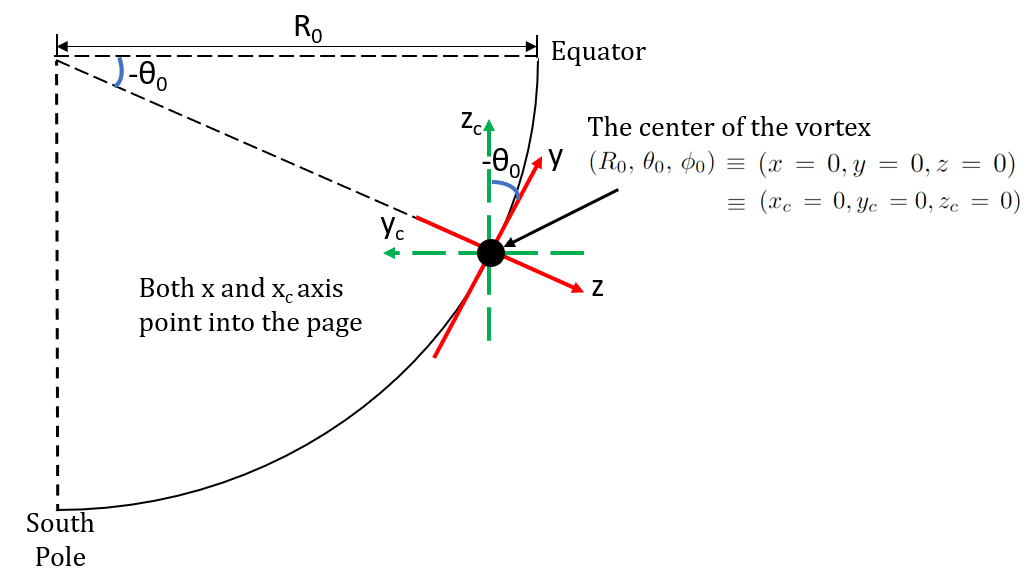}
    \caption{Two sets of  Cartesian coordinates. The thin quarter-circle indicates the cloud-top level of the planet. The South pole and Equator's locations are marked. The origin of both Cartesian coordinates (black dot) is at the center of the vortex ($r_0,\,\theta_0,\,\phi_0$). The angle shown is $-\theta_0$ because the GRS is in the southern hemisphere. 
    For the Cartesian coordinates $(x,\,y,\,z)$ in red, $\hat{\bf{x}}$ is eastward, $\hat{\bf{z}}$ is radially outward, and $\hat{\bf{y}}=\hat{\bf{z}}\cross\hat{\bf{x}}$ is along the local North direction. For the Cartesian coordinates $(x_c,\,y_c,\,z_c)$ in green, $\hat{\bf{x}}_c$ is eastward, $\hat{\bf{z}}_c$ is along the direction of the planetary rotational axis, and $\hat{\bf{y}}_c=\hat{\bf{z}}_c\cross\hat{\bf{x}}_c$. This illustration is chosen with $\theta_0 < 0$, which would be appropriate for the Great Red Spot of Jupiter, which lies in the southern hemisphere.}
    \label{fig:coordinate_conversion}
\end{figure}

 A commonly used coordinate system for studying planetary scale flows are spherical coordinates $(r,\theta,\phi)$, where r is the radius, $\theta$ the latitude, and $\phi$ the longitude. However, planetary vortices are local structures in planetary atmospheres. The characteristic horizontal length ${\cal L}$ of a typical planetary vortex is small compared to the planet's radius. The largest planetary vortex, the Great Red Spot (GRS) of Jupiter, spans $\sim11^\circ$ (longitude) $\cross \, 7.4^\circ$ (latitude), or  $13,000 \times 9,200$~km \citep{simon2018historical,wong2021evolution}, so ${\cal L} \simeq 13,000$~km. Because ${\cal L}$ is small compared to the mean Jovian radius $r_0 \simeq 72,000$ at the 1~bar level (near the tops of the visible clouds), we use a local Cartesian coordinate system (c.f., Pedlosky (1982) chapter 6). Let $(r_0, \theta_0, \phi_0)$ be the center of the vortex of interest, then the local Cartesian coordinates in Figure~\ref{fig:coordinate_conversion} are 
 \begin{eqnarray}
x&=&[r_0 \, \cos(\theta_0)] (\phi - \phi_0)  \, \biggl[1 + {\cal O}\bigl({\cal L}/r_0 \bigr)^2 \biggr]\label{eq:sph2cart_x}\\
y&=&r_0 \, (\theta-\theta_0) \,  \biggl[1 + {\cal O}\bigl({\cal L}/r_0 \bigr)^2 \biggr] \label{eq:sph2cart_y}\\
z&=&r -r_0  \, \biggl[ 1 + {\cal O}\bigl({\cal L}/r_0 \bigr)^2 \biggr] \label{eq:sph2cart_z}.
\end{eqnarray}
 Note $\hat{\bf{z}}$ is radially outward unit vector and is along the vertical direction in which the flow is stratified; $\hat{\bf{x}}$ is eastward; and $\hat{\bf{y}}=\hat{\bf{z}}\cross\hat{\bf{x}}$ is northward.  When computing the GRS at $\theta_0 \simeq -23^{\circ}$, we can ignore Jupiter's small oblateness and neglect the curvature correction terms because $({\cal L}/r_0)^2 \simeq 0.06$ in equations (\ref{eq:sph2cart_x}) --~(\ref{eq:sph2cart_z}). To exploit further approximations in the following sections, we shall also need to consider the horizontal length scale $L$, the characteristic distance associated with the horizontal derivative of the vortex. For example, the operator $(\partial^2 /\partial x^2 + \partial^2 /\partial y^2)$ acting on the velocity or pressure of a vortex will be of order $1/L^2$. For some vortices, $L$ can be considerably different from ${\cal L}$. For example, the GRS has a relatively quiet interior surrounded by a high-velocity collar of thickness $\sim2,000$~km, so although ${\cal L} \simeq 13,000$~km, $L \simeq 2000$~km.\footnote{The width of the high-velocity collar might be related to  Rossby deformation radius $L_R$. In two-dimensional, quasigeostrophic studies \citep{marcus1988numerical,dowling1989jupiter,shetty2010changes} have used a ``depth-averaged'' value of  $L_R\sim2,000$~km to try to qualitaveily simulate the velocity field of the GRS at cloud-top level. However, $L_R \equiv H_P \, \bar{N}/f$ where  $\bar{N}$ is the Brunt-Vaisala frequency, $H_P$ the local vertical pressure scale height, and $f$ the local value of the Coriolis parameter. Because $\bar{N}$ varies by three orders of magnitude over the height of the GRS, so does $L_R$. (See section \ref{sec:hydrostatic_fields_numerical})} 
 
 Local Cartesian coordinates are useful because much of the dynamics are controlled by gravity in the $z$ direction, which strongly stratifies the flow in that direction. However, Coriolis forces due to the planetary spin are also important, especially for low Rossby number flows. In the absence of baroclinicity and stratification, the vortices would obey the Taylor-Proudman theorem and be invariant along the direction parallel to the planet's spin axis. Therefore, it is convenient to introduce another local Cartesian coordinate system $(x_c,\, y_c,\, z_c)$, which has the same origin as the $(x,\,y,\,z)$ coordinates but is aligned with the spin of the planet (see Fig.~\ref{fig:coordinate_conversion}) such that $\hat{\bf z}_{\bf c}$ is in the direction of the planet's rotation vector, $x_c \equiv x$, and $y_c$ is orthogonal to $x_c$ and $z_c$ such that the right-hand rule is satisfied. To convert between the two local Cartesian coordinate systems:
\begin{eqnarray}
x&=&x_c\label{eq:cartc2cart_x}\\
y&=&y_c \, \sin\, \theta_0+z_c \, \cos \, \theta_0\label{eq:cartc2cart_y}\\
z&=&-y_c \, \cos \, \theta_0+z_c \, \sin \, \theta_0\label{eq:cartc2cart_z}\\
x_c&=&x\label{eq:cart2cartc_x}\\
y_c&=&y \, \sin \, \theta_0-z \, \cos \, \theta_0\label{eq:cart2cartc_y}\\
z_c&=&y \, \cos \, \theta_0+z \, \sin \, \theta_0. \label{eq:cart2cartc_z}
\end{eqnarray}
Note that the $(x, y, z)$ coordinates of the straight line parallel to the spin axis of the planet that passes through the point $(x', y', z')$ obeys
\begin{eqnarray}
x&=&x' \label{eq:obey1}\\
y&=&y' +(z-z') \, \cot \, \theta_0. \label{eq:obey2}
\end{eqnarray}
In other words, along the line described by equations~(\ref{eq:obey1}) --~(\ref{eq:obey2}), $x_c$ and $y_c$ are fixed, while $z_c$ varies.
We will need these relations to construct our initial conditions of the vortices.

\subsection{Anelastic Approximation and the Governing Equations}\label{sec:analytic_anelastic}
\subsubsection{Anelastic Approximation}\label{sec:analytic_anelastic_approximation}
We denote the solutions of the  steady hydrostatic (i.e., the velocity ${\bfv} = 0$) equations with overbars. The Euler equation and ideal gas equation become
\begin{eqnarray}
\dv{\bar{P}(z)}{z}&=&-\bar{\rho}(z) g\label{eq:hydrostatic_z}\\
\bar{P}(z)&=&R \, \bar{\rho}(z) \, \bar{T}(z), \label{eq:hydrostatic_ideal_gas}
\end{eqnarray}
where $P$ is pressure, $\rho$ is mass density, $T$ is temperature, $g$ is the acceleration of gravity pointing in the $-\hat{\bf z}$ direction, and $R$ is the ideal gas constant of the hydrogen-helium mixture of the planetary atmosphere of the interest, which for Jupiter 
at 1~bar is $R=3.60 \times 10^3$~m$^2$s$^{-2}$K$^{-1}$ \citep{de2019jupiter}. 
For the anelastic approximation, it is useful to introduce the potential temperature $\Theta$ with
\begin{eqnarray}
    \Theta&\equiv&T\left(\frac{P_{ref}}{P}\right)^{R/C_p}\label{eq:ptemp_def}\\
    \bar{\Theta}(z)&=&\bar{T}(z)\left(\frac{P_{ref}}{\bar{P}(z)}\right)^{R/C_p}, \label{eq:hydrostatic_ptemp_def}
\end{eqnarray}
where $P_{ref}$ is a reference temperature and $C_p$ is the heat capacity at constant pressure. To completely define the hydrostatic solution, an energy equation is needed. However, as will be discussed in $\S$~\ref{sec:new}, we replace that equation with the observed value of $\bar{T}(z)$. Using the observed $\bar{T}(z)$, all of the other barred thermodynamic quantities can computed with equations~(\ref{eq:hydrostatic_z}) --~(\ref{eq:hydrostatic_ptemp_def}). 

We decompose the thermodynamic variables as
\begin{eqnarray}
\tilde{\rho}(x,y,z,t)&=&\rho(x,y,z,t)-\bar{\rho}(z)\label{eq:rho_decompose}\\
\tilde{P}(x,y,z,t)&=&P(x,y,z,t)-\bar{P}(z)\label{eq:p_decompose}\\
\tilde{T}(x,y,z,t)&=&T(x,y,z,t)-\bar{T}(z)\label{eq:T_decompose}\\
\tilde{\Theta}(x,y,z,t)&=&\Theta(x,y,z,t)-\bar{\Theta}(z)\label{eq:ptemp_decompose}
\end{eqnarray}
The anelastic approximation requires that the
non-hydrostatic thermodynamic variables (denoted with the tildes) are much smaller their hydrostatic counterparts (denoted with the overbars) at each height $z$. In addition, the Mach number must be small.
These requirements are consistent with Jovian observations.

\subsection{Anelastic Equation of Motion}\label{sec:analytic_anelastic_equation}

There are different versions of the anelastic equations \citep{ogura1962scale,gough1969anelastic,bannon1996anelastic,brown2012energy,adrian1984scattering}, and we chose the one developed by \citet{bannon1996anelastic} because it conserves energy and we have found it to be robust in astrophysical calculations \citep{barranco2005three,marcus2015zombie,marcus2016zombie,barranco2018zombie}.\footnote{This equation set is very similar with the Lantz-Braginsky-Roberts (LBR) equations \citep{lantz1992dynamical,lantz1999anelastic,braginsky1995equations}. The only difference is that our thermal computational variable is potential temperature but the one in the LBR equation is entropy. \citet{brown2012energy} show that the LBR equations capture dynamics correctly in subadiabatic atmosphere because of energy conservation. Whereas some other versions of anelastic equations do not.} In the Cartesian coordinates $(x, y, z)$ shown in figure~\ref{fig:coordinate_conversion}, these equations are
\begin{eqnarray}
    0&=&\div{(\bar{\rho} \bfv)}\label{eq:anelastic_density}\\
    \pdv{\bfv}{t} + (\bfv\cdot\grad)\bfv-\frac{\tilde{\Theta}}{\bar{\Theta}}g\hat{z}+\grad{\left(\frac{\tilde{P}}{\bar{\rho}}\right)} &=& \bfv\cross\bff \nonumber\\
    &\simeq& f(v_y \, \sin \, \theta_0 - v_z \, \cos \, \theta_0) \, \hat{\bf x}  \nonumber \\
    &+& \beta \, y \, (v_y  + v_z \, \tan \, \theta_0) \, \hat{\bf x} \nonumber \\
    &-&v_x \, (f \sin \, \theta_0 + \beta y) \, \hat{\bf y} \nonumber \\
    &+&v_x \, (f \cos \, \theta_0 - \beta y \, \tan \, \theta_0) \, \hat{\bf z}  \label{eq:anelastic_momentum}\\
    \pdv{}{t}\,\left(\frac{\tilde{\Theta}}{\bar{\Theta}}\right)&=&-\frac{(\bfv\cdot\grad){\tilde{\Theta}}}{\bar{\Theta}}-\frac{\bar{N}^2}{g}v_z \nonumber \\ &=& -\frac{(\bfv_{\perp} \cdot\grad_{\perp}) {\tilde{\Theta}}}{\bar{\Theta}}-\frac{{N}^2}{g}v_z \label{eq:anelastic_energy} \\
    \frac{\tilde{\rho}}{\bar{\rho}}&=& \frac{C_v}{C_p} \, \frac{\tilde{P}}{\bar{P}} - \frac{\tilde{\Theta}}{\bar{\Theta}}, \label{eq:anelastic_ddensity}\\ 
    \frac{\tilde{T}}{\bar{T}}&=&\frac{\tilde{P}}{\bar{P}}-\frac{\tilde{\rho}}{\bar{\rho}}, \label{eq:anelastic_ideal_gas_law}
\end{eqnarray}
where $N$ is the Brunt-V\"{a}is\"{a}l\"{a} frequency 
\begin{equation}
{N}^2(x, y, z)\equiv\frac{g}{\bar{\Theta}}\frac{\partial {\Theta}}{\partial z}  \,\,\,\,\,\, {\rm and} \,\,\,\,\,  \bar{N}^2(z)\equiv\frac{g}{\bar{\Theta}}\dv{\bar{\Theta}}{z} = \frac{g}{\bar{T}(z)}\Big(\dv{\bar{T}(z)}{z}+\frac{g}{C_p}\Big), \label{eq:anelastic_N2}
\end{equation}
and where $\bff=f \, \hat{\bf{z}}_c$ is the full Coriolis vector, which is along the direction of $\hat{\bf{z}}_c$, or  $\bff = f \, \cos\theta \, \hat{\bf{y}}+f \, \sin\theta \, \hat{\bf{z}} \simeq (f \, \cos\theta_0 - \beta \, \tan \theta_0 \, y) \, \hat{\bf{y}}+(f \, \sin\theta_0 + \beta \, y) \, \hat{\bf{z}}$, 
where $\beta \equiv (f \, \cos \theta_0)/r_0$. 
In writing the gravitational acceleration as $- g \, \hat{\bf z}$, rather than a vector in the spherical radial direction, we have ignored the small curvature correction terms. We also ignored the higher-order curvature terms in the Coriolis vector and the centrifugal force \citep{pedlosky1987geophysical}. 

In equation~\eqref{eq:anelastic_momentum}, the viscous term is dropped. In equation \eqref{eq:anelastic_energy}, the thermal conduction and radiative transfer terms are not considered. As the system of interest has a high Reynolds number, the dissipation length scale is much smaller than the numerical resolution, and we use hyperviscosity and hyperdiffusivity to perform numerical dissipation at the smallest resolvable scales. Dropping the viscous, thermal conduction, and radiative transfer terms means that the convective velocities below the top of the underlying convection zone are not computed. However, the near-adiabatic temperatures of the underlying convection zone are used in these calculations, and the Brunt-V\"{a}is\"{a}l\"{a} frequency smoothly goes from the observed large value at the top of the computational domain down to zero at the top of the convective zone. (See $\S$~\ref{sec:analytic_hydrostatic_equlibrium_general} and Figure~2.) Moreover, although we have omitted the convective velocities within the convection zone itself, we do accurately compute the intermittent convective velocities above the top of the convection zone when and where the flow becomes locally superadiabatic and unstable to local convection -- see $\S$~\ref{sec:numerics_convection}.   
We do not believe that our calculations are harmed by the lack of convective velocities below the top of the convection zone because the bottoms of our computed vortices lie well above the convection zone and never penetrate down to the top of the convective zone.
The details of our numerical solver and boundary condition are discussed in the Appendix.

To compute our numerical solutions using an initial-value code in $\S$~\ref{sec:numerical}, we numerically solve (\ref{eq:anelastic_density}) --~(\ref{eq:anelastic_ideal_gas_law}). However in $\S$~\ref{sec:IC} where we compute initial conditions of the vortices that are ``close to'' equilibrium flows, it is more  useful to replace the vector momentum equation~(\ref{eq:anelastic_momentum}) with an equivalent set of three scalar equations: the vertical component of the vorticity equation; the horizontal divergence of the horizontal component of the momentum equation; and the vertical component of the momentum equation. These equations are:
\begin{eqnarray}
\pdv{(\omega_z + \beta \, y)}{t}&=&-(\bfv_\perp\cdot\nabla_\perp)\, (\omega_z + \beta \, y) -v_z\pdv{\omega_z}{z}+(\bfome_\perp\cdot\nabla_\perp)v_z \nonumber \\
&-&\omega_z \, (\nabla_\perp\cdot\bfv_\perp) -(f \, \sin\theta_0 + \beta \, y)  \, (\nabla_\perp\cdot\bfv_\perp) \nonumber \\
&+&(f \, \cos \theta_0 - \beta \, y \, \tan \theta_0)\, \pdv{v_z}{y} - \beta \, v_z \, \tan \theta_0
\label{eq:domegaz_dt} \\
{{\partial (\nabla_{\perp} \cdot {\bfv}_{\perp})}\over{\partial t}} &=&-\nabla_\perp\cdot \bigl[(\bfv\cdot\nabla)\bfv_{\perp} + \bfv\cross\bff \bigr]
- \nabla^2_{\perp}
\biggl( \frac{\tilde{P}}{\bar{\rho}} \biggr) \label{eq:second} \\
{{\partial v_z}\over{\partial t}}  &=& - ({\bfv} \cdot \nabla) \, v_z + g \frac{\tilde{\Theta}}{\bar{\Theta}} -\pdv{ }{z}\,\left(\frac{\tilde{P}}{\bar{\rho}}\right)  +f \, v_x \, \cos \theta, \label{eq:third}
\end{eqnarray}
where $\bfome$ is the vorticity vector, $\bfv_\perp$ is the  horizontal velocity, $\bfome_\perp$ is the horizontal vorticity,  $\nabla_\perp \equiv \hat{x}\frac{\partial}{\partial x}+\hat{y}\frac{\partial}{\partial y}$, $\nabla_\perp^2=\pdv[2]{ }{x}\,+\pdv[2]{ }{y}$ is the horizontal Laplacian operator, and  $\nabla_\perp\cdot\mbox{\boldmath $V$} \equiv \pdv{}{x}\,(\mbox{\boldmath $V$}\cdot\hat{\bf{x}})\,+\pdv{ }{y}\,(\mbox{\boldmath $V$}\cdot\hat{\bf{y}})$ is the horizontal divergence of any given vector $\mbox{\boldmath $V$}$. 

\section{Choice of Hydrostatic Fields} \label{sec:new}
\subsection{Our Non-Use of the Hydrostatic Energy Equation}\label{sec:analytic_hydrostatic_equlibrium_general}
The reason we use the observed $\bar{T}(z)$, rather than solve the energy equation to find it, is because the former is relatively easy to obtain and the latter is difficult to solve. The equation governing the energy or temperature of the flow is
\begin{eqnarray}
{{\partial T}\over{\partial t}} &+& ({\bfv} \cdot \nabla) T =  \frac{\nabla \cdot( k \nabla {T})}{C_v \rho} - \frac{P \, (\nabla \cdot {\bfv})}{C_v \rho} \nonumber \\  
&+& Source/Sink\:terms\:due\:to\:radiative\:transfer \: and \:  viscous \: heating, \label{eq:energy_equation_background}
\end{eqnarray}
where $k$ is the coefficient of heat conduction and $C_v$ is the heat capacity at constant volume of the fluid. Because the flow is optically thin at heights above the visible cloud tops, the radiative heat transfer responsible for heating and cooling is complex (especially in regions with different cloud decks where there are phase changes). 

A second complication of this equation is that, like equation~(\ref{eq:anelastic_energy}), it will have convective velocities in any region where $N(x, y, z, t)^2 < 0$. To truly simulate Jovian flows, we would need to use equation~(\ref{eq:energy_equation_background}) with correct parameter values of $k(x, y, z)$, the cloud compositions, and the radiation physics such that the top of the convection zone would be at 10~bar, with large convective velocities in the convection zone. Those velocities would mix the potential temperature, making the time-averaged temperature within the convection zone approximately adiabatic with  $\bar{N} \simeq  0$.\footnote{Local, intermittent convection has been inferred in Jupiter at heights above 10~bar, but it is not generally believed that the flow is fully convective at heights above 10~bar.}  While experience has shown us that we can compute convection with the anelastic equations \citep{marcus1978nonlinear}, and \citet{yadav2020deep} and other authors have done so, we believe that the dynamics of 3-dimensional vortices is sufficiently complex that this first study should be carried out with the vortices in a stable rather than turbulently convective ambient atmosphere. Also, like \citet{morales2013jupiter}, we wish to explore freely evolving Jovian vortices, rather than those forced by convection. For these reasons, we artificially suppress global convection (but allow local convection) in this study by choosing a $\bar{T}(z)$ that agrees with observations and is weakly sub-adiabatic (i.e., with $N = 0^+$) beneath the top of the convective zone at 10~bar. Thus, $\bar{T}(z)$ is close to its actual value in the convection zone, but the fluid deeper than 10~bar is not filled with turbulent convection.\footnote{This replacement is what was done in the early days of calculating stellar structure \citep{schwarzschild2015structure}. Typically, one solved the steady-state energy equation along with the other hydrostatic equations, and in any region in which the vertical gradient of the temperature made the flow unstable to convection, the temperature was replaced with an adiabat such that $\bar{N}=0$.} Our underlying philosophy is that some of the long-lived vortices computed with this approximation might be destroyed or modified by convection. However, our belief is that there are no families of unforced vortices that could be computed with a true convective zone that would not have a counterpart using this approximation. We suspect that vortices computed with this approximation that extend into the true convective zone would be destroyed if convection were actually present, or at least, have their bottom parts that extend into the convective zone truncated.  (See our Discussion.)

\subsection{Constructing Hydrostatic Fields Based on Observations}\label{sec:hydrostatic_fields_numerical}
\begin{figure}
\begin{subfigure}{0.49\textwidth}
\includegraphics[width=66mm]{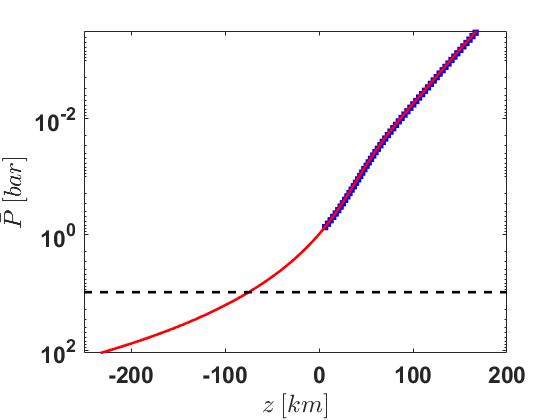}
    \caption{$z\:\mathrm{[km]}$}
    \label{fig:thermal_bk_z}
\end{subfigure}
\begin{subfigure}{0.49\textwidth}
\includegraphics[width=66mm]{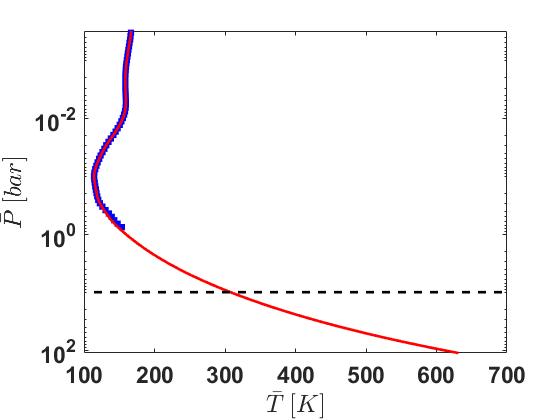}
    \caption{$\bar{T}\:\mathrm{[K]}$}
    \label{fig:thermal_bk_T}
\end{subfigure}\\
\begin{subfigure}{0.49\textwidth}
\includegraphics[width=66mm]{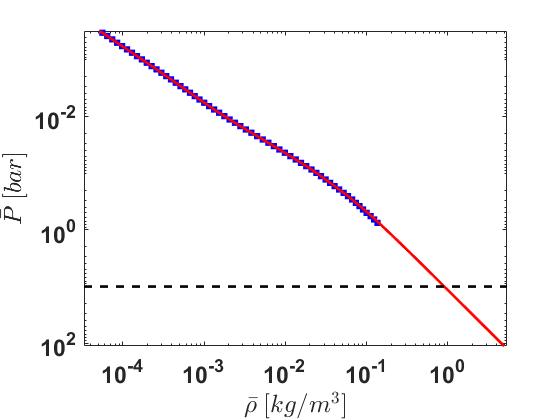}
    \caption{$\bar{\rho}\:\mathrm{[kg/m^3]}$}
    \label{fig:thermal_bk_rho}
\end{subfigure}\begin{subfigure}{0.49\textwidth}
\includegraphics[width=66mm]{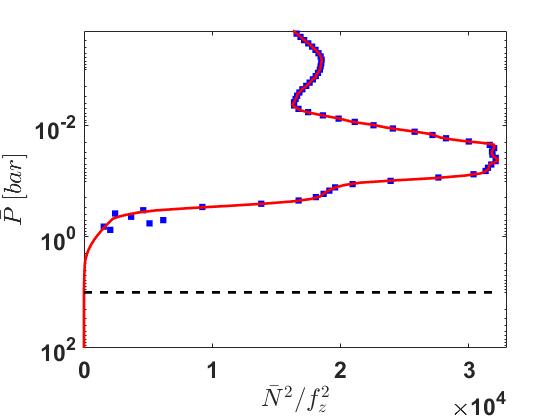}
    \caption{$\bar{N}^2\:\mathrm{[s^{-2}]}$}
    \label{fig:thermal_bk_N2}
\end{subfigure}
\caption{Comparison of the vertical profiles of the background thermodynamic functions based on the observed $\bar{T}(\bar{P})$ (blue dots) and the smoothed functions (red curves). The black dashed line is the top of the convection zone at 10~bar. We define $z=0$ as the plane where $\bar{P}=1$~bar, near the top of the visible clouds. \label{fig:thermal_bk}}
\end{figure}

The thermal structure of Jovian atmosphere is crucial for the existence of Jovian vortices. However, directly simulating the radiative transfer and convective processes that determine the Jovian thermal structure is complex, as described in \S \ref{sec:analytic_hydrostatic_equlibrium_general}. 
Therefore for simplicity, we use the observed zonal-averaged values of $\bar{T}(\bar{P})$ at the GRS latitude $\theta_0=23^\circ$S, as specified by \citet{de2019jupiter,moeckel2022ammonia} along with equations~(\ref{eq:hydrostatic_z}) and~(\ref{eq:hydrostatic_ideal_gas}) to obtain first guesses of 
$\bar{P}(z)$, $\bar{P}(\bar{T})$, and $\bar{P}(\bar{\rho})$ as shown by the blue dots in figure~\ref{fig:thermal_bk}. Although the temperature data appears to be smooth, when we differentiate it to compute $\bar{N}(z)$ with equation~(\ref{eq:anelastic_N2}), we find it is not smooth, especially at heights deeper than $\sim200$~mbar (See the blue dots in figure~\ref{fig:thermal_bk_N2}). A non-smooth $\bar{N}(z)$ has the potential of producing artificial convective instabilities and internal gravity waves in our calculations. In addition, the measured values of temperature only extend to a depth of  $\sim$~1~bar. Therefore we have taken the observed temperature measurements and change them in three ways: (1) at heights above $800$~mbar we replace the observed $\bar{N}^2(z)$ (blue dots in figure~\ref{fig:thermal_bk_N2}) with the locally smoothed red curve; (2) we define the top of the convective zone to be at 10~bar, so at heights deeper than 10~bar, we set $\bar{N} = 0^+$; and (3) we
construct a smooth monotonic curve (red) to extrapolate $\bar{N}$ between $800$~mbar and 10~bar. We then use equations~(\ref{eq:hydrostatic_z}), (\ref{eq:hydrostatic_ideal_gas}), (\ref{eq:hydrostatic_ptemp_def}), and~(\ref{eq:anelastic_N2}) with the smooth $\bar{N}$ in figure~\ref{fig:thermal_bk_N2}, to compute the smoothed (red) functions $\bar{P}(z)$, $\bar{P}(\bar{T})$, and  $\bar{P}(\bar{\rho})$  in figure~\ref{fig:thermal_bk}abc.
The smoothed functions are not very different from the original observational data. 

\section{Choice of Zonal Flow} \label{sec:zonal_flow}
\begin{figure}
    \centering
    \includegraphics[trim={2cm 0 0 0},clip,width=135mm]{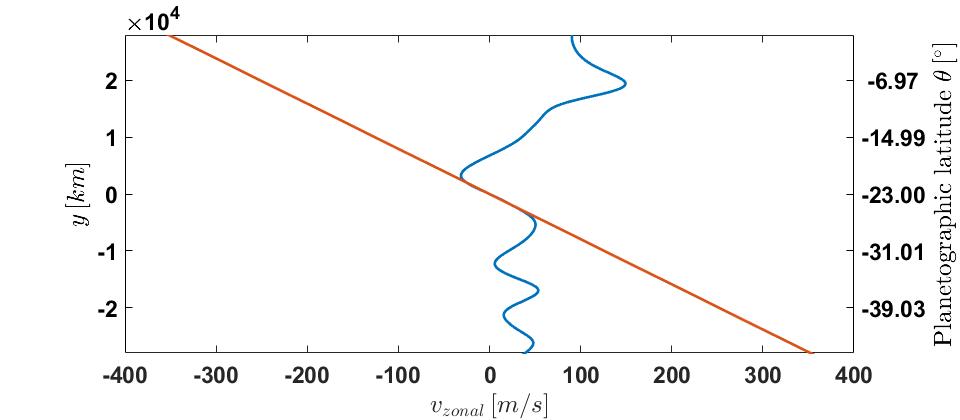}
    \caption{The observed Jovian zonal velocity (blue) at a radius of the planet $R$ corresponding to a pressure of approximately 1~bar vs a constant-shear zonal velocity (red) with $v_{zonal} = - \sigma y$, where $\sigma = -9.1\cross 10^{-6}\,\mathrm{s}^{-1}$. The velocity is shown as a function of planetographic latitude $\theta$ and of the north-south length $y$, with  $y=0$ chosen to be the approximate center of the GRS.}
    \label{fig:vzonal}
\end{figure}

The Jovian east-west zonal flow $v_{zonal}(y, z) \, \hat{\bf x}$, far from the large vortices is, by definition,  averaged over longitude, and is therefore independent of $x$. At the height of the visible cloud tops, it has been measured repeatedly \citep{limaye1986jupiter,porco2003cassini,asay2011changes,tollefson2017changes} and observed to be 
nearly independent of time. The observations are shown as a function of latitude $\theta$ or $y$ in figure~\ref{fig:vzonal}. For the calculations here, we need to know $v_{zonal}$ for all values of $z$ that contain a Jovian vortex. The only direct measurement of $v_{zonal}$ at elevations other than the cloud tops was from the {\it Galileo} probe \citep{atkinson1998galileo}, which measured the velocity at only one location, and that result is controversial because the location was a ``hot spot'' with anomalous properties \citep{marcus2019equatorial}. Furthermore, because the probe entered at latitude 7.3$^{\circ}$N jovigraphic, it measured velocities along a descent path more aligned with the $y_c$ axis than the $z_c$ axis. For low Rossby number flows, the thermal wind equation \citep{pedlosky1987geophysical}, or the equivalent density wind equation \citep{kaspi2020comparison}, can be used to determine the vertical derivative of the zonal flow. However, the use of those equations requires accurate thermal $\tilde{T}$ or density $\tilde{\rho}$  data as a function of $(y,\,z)$. The temperature and density measurements are sufficiently uncertain that the results are ambiguous \citep{fletcher2016mid,fletcher2021jupiter}. Observations based on {\it Juno} gravity measurements \citep{kaspi2020comparison} suggest that the zonal flow beneath the cloud-tops depends primarily on $y_c$ and has an exponential decay in the $z$ direction with an e-folding length of $1471$~km, which is much greater than the vertical size, $\sim 200$~km, of the Jovian vortices studied here. 

Determining the exact vertical structure of the zonal flow is not the focus of this study. Therefore for simplicity, we choose a zonal flow that is independent of $z_{c}$, satisfies the steady anelastic equations, and agrees with the observations at the cloud tops. This simple form of zonal flow is consistent with the {\it Juno} gravity measurement \citep{kaspi2020comparison}, which shows that the e-folding depth of Jovian zonal flow is $\sim1471$~km, which is much bigger than the vertical size of the domain in this study. It is also consistent with the recent James Webb Space Telescope (JWST) zonal wind measurement \citep{hueso2023intense}, which shows that the zonal wind near the top of our vortex ($\sim140$~mbar) is similar to the cloud-top measurement, at the latitude of our interest ($23^\circ$S).

{\it Any} zonal flow ${\bfv} = v_{zonal}(y, \, z) \, \hat{\bf x}$ is a solution to the steady anelastic equations subject to the  requirements that 
\begin{eqnarray}
    {{\tilde{P}(x, y, z)}\over{\bar{\rho}(z)}} &=& - \int_{0}^y \, dy' \, v_{zonal}(y', z) \, (f \, \sin \, \theta_0 + \beta \, y') \, + {{\tilde{P}(x, 0, z)}\over{\bar{\rho}(z)}} \label{eq:zoneym} \\
    {{\tilde{\Theta}}\over{\bar{\Theta}}} &=& {{1}\over{g}} \,\Bigg[  {{\partial}\over{\partial z}}\Biggl({{\tilde{P}}\over{\bar{P}}} \Biggr)  - v_{zonal} \, (f \, \cos \, \theta_0 -  \beta \, y \, \tan \, \theta_0)\Bigg] \label{eq:zoneyn}
\end{eqnarray}
Therefore, the only constraint on our choice of  zonal flow is
that
\begin{eqnarray}
v_{zonal}(y,z) = {\cal F}(y_c) \equiv {\cal F}(y \, \sin \theta_0 - z \, \cos \, \theta_0),
\end{eqnarray}
where ${\cal F}(y \, \sin \theta_0 - z \, \cos \, \theta_0)$ matches the Jovian observations in figure~\ref{fig:vzonal} at $z=0$ (the value of $z$ near the visible cloud-tops). Two different zonal flows $v_{zonal}(y_{c})$ are used in this study. One corresponds to the blue curve in figure~\ref{fig:vzonal}, which is the observed Jovian zonal velocity, and the other is an approximation (red curve) that uses a zonal velocity with constant shear.\footnote{The observed zonal velocity is smoothed to avoid numerical instabilities. Both choices of $v_{zonal}(y_{c})$ are artificially made periodic near the y directional boundaries. See the Appendix.}  
We include a study of vortices with the constant-shear zonal (red line in Figure~3) flow because we want to determine whether the local maximum and minimum of the zonal flow have a strong effect on the ability of a vortex to be hollow (as proposed by \citet{shetty2007interaction}) and whether locations where the Rossby Mach number becomes supersonic control the flow's stability (as proposed by \citet{dowling2014saturn,dowling2019jets,dowling2020jupiter,afanasyev2022evolution}). We note that we carried out numerous simulations in which the initial condition consisted of a zonal flow, finite-amplitude ``noise'', and no vortices. In no case did the noise de-stabilize the zonal flow, so we argue that the zonal flows used in these computations and these boundary conditions are linearly stable and also stable to a variety of finite-amplitude perturbations. We note that if the zonal flows of Jupiter extend into the convective zone, as many researchers believe, then the stability of the actual Jovian zonal flows cannot be determined by our calculations. 

\section{Initial Condition Set Up}\label{sec:IC}
\subsection{Strategy}\label{sec:IC_beginning}

The unforced and undissipated equations that govern the flow do not have a unique solution. For example, $\tilde{P} = \tilde{\rho} = \tilde{\Theta} = \tilde{T} = {\bfv} =0$ is a solution with no zones and no vortices. Purely zonal flows $\bfv_{zonal}(y,z) \equiv v_{zonal}(y, z) \, \hat{\bf x}$ with no vortices also exist with the zonal vorticity equal to $\bfome_{zonal} \equiv \curl  \bfv_{zonal}$, 
and with a zonal pressure $\tilde{P}_{zonal}$, zonal temperature $\tilde{T}_{zonal}$, and zonal potential temperature $\tilde{\Theta}_{zonal}$ determined by  equations~(\ref{eq:zoneym}) and~(\ref{eq:zoneyn}) with $\bfv=\bfv_{zonal}(y,z)$.
In addition, there are families of solutions that have vortices embedded in the zonal flow. 
For these flow, it is useful to decompose the velocity into contributions due to the vortex as $\bfv_{vortex}\equiv \bfv-\bfv_{zonal}$, with  similar decompositions for pressure, temperature, and potential temperature. 
For the vertical vorticity, the contribution due to the vortex is  $\omega_{z,vortex}  \equiv (\bfome  - \bfome_{zonal}) \cdot \hat{\bf z}$ where $\bfome_{zonal} \cdot \hat{\bf z} \equiv - \partial v_{zonal}/\partial y$.
We also define $\omega_{z,vortex,0}(x,y) \equiv \omega_{z,vortex}(x,y, z=0)$. Throughout the remainder of this paper, we define $z=0$ to be 1~bar, which we take to  be the nominal  height of the visible Jovian cloud tops where the horizontal velocity is observed.\footnote{Some observers believe the cloud tops can be as high as 500~mbar.}

\subsection{Stacked Models}\label{sec:stacked_model}

Here we show how to construct an initial vortex that is ``close to'' an equilibrium solution of the governing equations of motion and that ``looks like'' a Jovian vortex because its $\omega_{z,vortex,0}(x,y)$ is in agreement with Jovian observations at the cloud-top level. To find these near-equilibria, we note that the vertical velocities $v_z$ of planetary vortices are small compared to their horizontal velocities and difficult to measure. Many studies assume $v_z=0$ and the quasigeostrophic equation \citep{marcus2000vortex,shetty2010changes,marcus2011jupiter} or shallow water equation \citep{dowling1989jupiter,cho1996emergence,stegner2000numerical,garcia2017shallow,li2020modeling} to look at vortex dynamics. 
To create near-equilibria, we assume $v_z=0$ and that the flow is steady in time. With these assumptions, equations~(\ref{eq:anelastic_density}), (\ref{eq:domegaz_dt}) --~(\ref{eq:third}), and (\ref{eq:anelastic_energy}) become
\begin{eqnarray}
\nabla_{\perp} \cdot {\bfv}_{\perp} &=& 0 \label{eq:a} \\
(\bfv_\perp\cdot\nabla_\perp)\, (\omega_z + \beta \, y) &=& 0 \label{eq:b} \\
\nabla^2_{\perp}
\biggl( \frac{\tilde{P}}{\bar{\rho}} \biggr) &=& \nabla_\perp\cdot \bigl[-(\bfv_{\perp}\cdot\nabla_{\perp})\bfv_{\perp}\bigr] + (f \, \sin \, \theta_0 + \beta \, y) \, \omega_z - \beta \, v_x \label{eq:c} \\
g \frac{\tilde{\Theta}}{\bar{\Theta}}  &=& \pdv{ }{z}\,\left(\frac{\tilde{P}}{\bar{\rho}}\right)  -v_x\, (f \, \cos \, \theta_0 -  \beta \, y \, \tan \, \theta_0) \label{eq:d} \\
(\bfv_\perp\cdot\nabla_\perp)\, \tilde{\Theta} &=& 0 \label{eq:e}
\end{eqnarray}
These equations, along with
equations~(\ref{eq:anelastic_ddensity}) --~(\ref{eq:anelastic_ideal_gas_law}), are equivalent to the governing equations of motion \eqref{eq:anelastic_density} to \eqref{eq:anelastic_ideal_gas_law} for a steady flow with $v_z=0$.
To construct an approximate numerical solution, we begin by solving equations~(\ref{eq:a}) and~(\ref{eq:b}) for ${\bfv}_{\perp}(x, \, y, \, z=z')$ for each collocation point $z'$. We can do this because equations~(\ref{eq:a}) and~(\ref{eq:b}) at one value of $z$ are decoupled from those at other values of $z$. Equations~(\ref{eq:a}) and~{(\ref{eq:b})} are equivalent to the steady, Euler equations for a 2-dimensional, incompressible velocity for a constant density fluid, and there are many ways to solve them. For example, contour dynamics \citep{shetty2006modeling} allows the computation of steady equilibrium flows even if they are unstable. The most popular method is to use solve the 2-dimensional Euler equation with an initial-value solver and let the flow come to a steady state.  In general if the zonal flow $v_{zonal}(y, z) \, \hat{\bf x}$ is given, then a multi-parameter family of steady, or slowly drifting vortices,  with $\omega_{vortex}$
embedded in the zonal flow can be found. Once a solution is found at every collocation point in $z$, the 2-dimensional solutions can be stacked (as described in the next section) on top of each other to create a steady 3-dimensional solution to equations~(\ref{eq:a}) and~(\ref{eq:b}).\footnote{If the initial 2-dimensional vortices at different heights $z$ drift at different speeds, it does not appear to be a problem. For all cases that we have examined so far, the initial vortex (which is not designed to be a solution to the full equations, but only to be close to an attracting solution) comes to a  statistically-steady vortex when observed in some Galilean frame moving in the $x$ direction.}

After finding ${\bfv}_{\perp}(x, y, z)$, we substitute it into right side of equation~(\ref{eq:c}) and solve for $\tilde{P}(x, y, z)$. We then use equation~(\ref{eq:d}) with $\tilde{P}(x, y, z)$ to find $\tilde{\Theta}(x, y, z)$. We then use $\tilde{\Theta}(x, y, z)$ in equations~(\ref{eq:anelastic_ddensity}) --~(\ref{eq:anelastic_ideal_gas_law}) to find $\tilde{\rho}(x, y, z)$ and $\tilde{T}(x, y, z)$. The only governing equation that is unsatisfied is (\ref{eq:anelastic_energy}), which is why this is only an approximate solution. In $\S$~\ref{sec:numerical} we show numerically that with some judicious choices, this approximate solution is close to an equilibrium solution and explain why. 

For the case of a vortex embedded in a uniform zonal shear flow and  $\beta \equiv 0$, \citet{moore1971structure} found a family 
of analytic, stable 2-dimensional solutions to equations~(\ref{eq:a}) --~(\ref{eq:c}) consisting of an elliptical vortex with uniform vertical vorticity $\tilde\omega_z=\omega_0$ inside the vortex. For $\bfv_{zonal}=-\sigma y\, \hat{\bf{x}}$, they found
\begin{equation}
\sigma/\omega_0 = \frac{\epsilon^2-\epsilon}{\epsilon+1}, \label{eq:ms}
\end{equation}
where $\epsilon$ is the horizontal aspect ratio (length of of the ellipse's major diameter in $x$ to its minor diameter in $y$). As $\sigma$ increases, zonal flow's shear stretches the vortex in the $x$ directions and $\epsilon$ increases. Note that Moore-Saffman vortices, like all steady, inviscid, constant-density,  2D solutions to Euler's equation, streamlines and vorticity contours that are coincident, so $(\bfv\cdot\nabla\omega_z)=0$. In the limit of $\omega_0\rightarrow0$, $\epsilon\rightarrow\infty$, which is the case where velocity induced by the vortex is negligible. The flow streamlines are strictly zonal. 
One of the properties of Moore-Saffman vortices that we shall exploit later in this section is that, for constant $\sigma$ and $\omega_0$, the ellipticity is determined by equation~(\ref{eq:ms}), but the diameter of the vortex is a free parameter. Therefore, regardless of the size of the vortex in equation~(\ref{eq:ms}), it is a solution to equations~(\ref{eq:a}) --~(\ref{eq:c}) with $\beta=0$.

Calculations of Moore-Saffman vortices in 2-dimensions show they are robust, so they are 
therefore potentially useful building blocks of initial 3-dimensional vortices 
even though they have $\beta \equiv 0$ and even though their vertical vorticities $\omega_{z,vortex}$ are uniform and do not look like the GRS, which has a pronounced local minimum of $\omega_{z,vortex}$ at or near its center (i.e., the GRS is ``hollow''). However, we do make one change to the Moore-Saffman vortices when we use them to build a non-equilibrium initial vortex. 
We smooth the outer edge of the vortex and include an annular region of opposite-signed $\omega_{z,vortex}$ at the outer edge, to create shielded vortices. Observations at the cloud-top level show that the Jovian vortices are shielded \citep{shetty2010changes,wong2021evolution,grassi2018first,scarica2022stability,choi2007velocity} and that the horizontal thickness of the GRS's opposite-signed annular shield is $\sim$~2000~km.
To create our Moore-Saffman-like vortices at the cloud-top level, we define the following elliptical coordinates for convenience.
\begin{eqnarray}
   \eta&=&\sqrt{\Bigl(\frac{x-x_0}{\epsilon}\Bigr)^2+(y-y_0)^2}\\
   \chi&=&\tan^{-1}\biggl[\frac{\epsilon(y-y_0)}{x-x_0}\biggr]
\end{eqnarray}
where $(x_0, y_0)$ is the center of the vortex. We define the 
velocity field  of the vortex as 
\begin{eqnarray}
   v_{vortex}(x,y)&=&
   \begin{cases}
   \frac{\epsilon}{1+\epsilon} \, \omega_0 \, \eta & (\eta<R_v) \label{eq:begin}\\
   \frac{\epsilon}{1+\epsilon} \, \omega_0 \, \frac{R_v^2} {\eta} \, e^{\frac{-\eta+R_v}{L_r}} & (\eta > R_v)
   \end{cases}\\
   \bfv_{vortex}(x,y)&=&-v_{vortex} \, (\sin \, \chi) \, \hat{\bf x} +
   v_{vortex} \, (\cos \, \chi) \, \hat{\bf y} \\
   \omega_{z,vortex}(x,y)&=&\pdv{}{x}\,v_{y,vortex}-\pdv{}{y}\,v_{x,vortex}, \label{eq:end}
\end{eqnarray}
where $R_v$ is the elliptical radius of the vortex (the locus where the outer annular, opposite-signed, shield begins), and $L_r$ is the characteristic thickness of the annulus. We use $L_r=2,000\text{ km}$ throughout this study. The value of $\omega_0$ is determined by $\epsilon$ via equation \eqref{eq:ms}, and $\omega_{z,vortex}(x_0,y_0)=\omega_0$. In the limit $L_r \rightarrow \infty$, the vortex given by equations~(\ref{eq:begin}) --~(\ref{eq:end}) is an exact Moore-Saffman vortex with vorticity $\omega_0$. 
Of course, neither the 3-dimensional initial vortex constructed from Moore-Saffman vortices or from the vortices in equations~(\ref{eq:begin}) --~(\ref{eq:end}) are exact equilibria of equations~(\ref{eq:a}) --~(\ref{eq:c}), but we shall show that they work well enough.

\subsection{Projection Methods}\label{sec:IC_method}

\begin{figure}
\begin{subfigure}{0.49\textwidth}
\includegraphics[width=66mm]{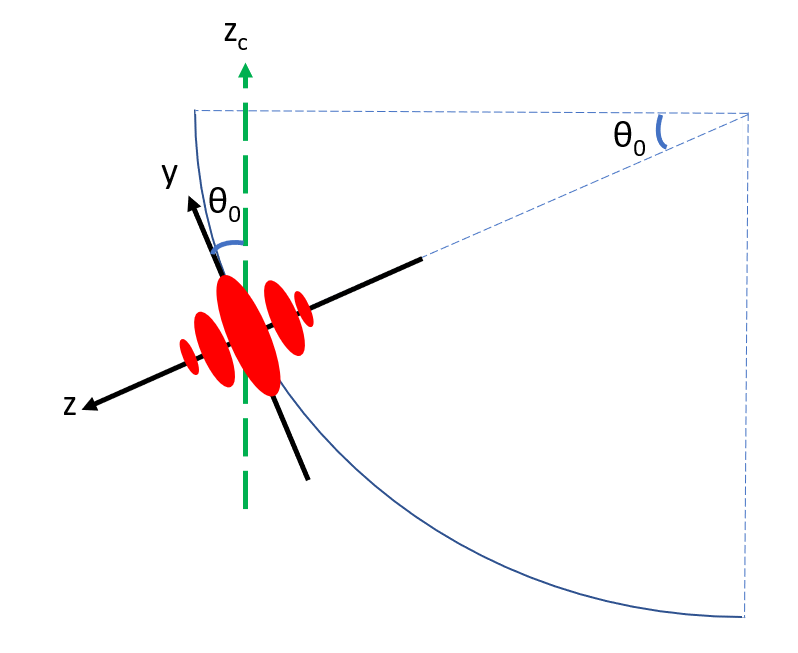}
    \caption{a vortex stacked along the radial axis $z$}
    \label{fig:wz_stacked_rad}
\end{subfigure}
\begin{subfigure}{0.49\textwidth}
\includegraphics[width=66mm]{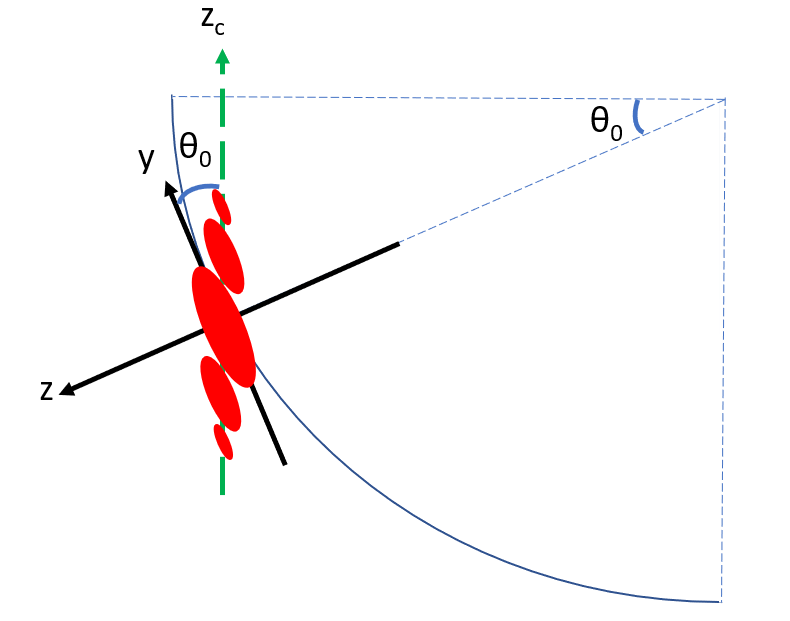}
    \caption{a vortex stacked along the rotational axis $z_c$}
    \label{fig:wz_stacked_rot}
\end{subfigure}
\caption{Two examples of 3-dimensional vortices made from stacked 2-dimensional vortices (red ellipses). The local Cartesian axes with $z$ oriented in the opposite direction as gravity are in black. The planetary-rotational axis $z_{c}$ is in green. $\theta_0$ is the latitude at the center of the vortex. Panel (a) is a vortex stacked along $z$. Panel (b) is a vortex stacked along $z_c$. Note that both panels show that the 2-dimensional vortices lie in $(x, y)$ planes and not $(x_c, y_c)$ planes. The vorticity of each 2-dimensional vortex is parallel to the $z$ axis, not the $z_c$ axis. For Jovian vortices, where the ratios of their vertical thicknesses to horizontal diameters are $\sim 0.01$, the figure is highly exaggerated. It would be difficult to directly observe the differences between the vortices in the two panels, if plotted with the true horizontal-to-vertical aspect ratio. \label{fig:wz_stacked}}
\end{figure}

Two-dimensional vortices that satisfy equations~(\ref{eq:a}) and~(\ref{eq:b}) can be stacked along any arbitrary direction and satisfy the 3-dimensional governing equations~(\ref{eq:a}) --~(\ref{eq:d}) and~(\ref{eq:anelastic_ddensity}) --~(\ref{eq:anelastic_ideal_gas_law}) 
(but not equation (\ref{eq:e})). Two evident directions along which to have the centers of each 2-dimensional vortex lie are the $z$ axis as in figure~\ref{fig:wz_stacked_rad} or the $z_c$ axis, parallel to the planet's rotation axis, as in figure~\ref{fig:wz_stacked_rot}. One would choose the former direction if the Rossby number were large so that the stratification was more important than the Coriolis forces (like a tornado), and choose the latter orientation if vice versa. For a low-Rossby-number barotropic flow, the Taylor-Proudman theorem would make the flow independent of $z_{c}$ and the vortex would be aligned along the $z_c$ axis. In either case, the strong stratification of the Jovian atmosphere makes the flow baroclinic, so that regardless of the orientation of the stacking, to satisfy equations~(\ref{eq:a}) --~(\ref{eq:d}) and~(\ref{eq:anelastic_ddensity}) --~(\ref{eq:anelastic_ideal_gas_law}), all of the 2-dimensional vortices lie in $x$--$y$ planes rather than $x_c$--$y_c$ planes, and the flows have $v_z = 0$, rather than $v_{z_c} = 0$. See figure~\ref{fig:wz_stacked}.

Although we shall show in the next section that a 3-dimensional vortex is well-approximated by a stack of 2-dimensional vortices with $v_z=0$, constructing 3-dimensional vortices from observations is challenging because currently the 2-dimensional velocity can only be measured at the level of the visible cloud tops. The velocity elsewhere has little observational constraint. 
For this reason at height $z=0$, we define a {\it reference} horizontal velocity $\bfv_{vortex,0}(x, y)$ and vertical vorticity  ${\omega}_{z,vortex,0}(x, y)$. In future studies, we shall set ${\omega}_{z,vortex,0}(x, y)$ equal to the observed values of the Jovian vortex we wish to study. Here, we forego observational velocities, and instead use the modified Moore-Saffman velocity in equations~(\ref{eq:begin}) --~(\ref{eq:end})  because in this study we are more concerned with finding solutions to the full 3-dimensional equations of motion where a known and controllable reference $\omega_{z,vortex,0}(x,y)$ is given. We are particularly interested in whether the vortex's final quasi-equilibrium $\bfv_{vortex}(x, y, z=0)$ is approximately the same as its initial reference $\bfv_{vortex,0}(x, y)$. In using equations~(\ref{eq:begin}) --~(\ref{eq:end}), the value of $\omega_0$ is slaved to the value of $\epsilon$ by equation~(\ref{eq:ms}) and the value of $\sigma$. In this paper we use $\sigma = -9.1\cross 10^{-6}\,\mathrm{s}^{-1}$ when calculating $\omega_0$ for all of the simulations, regardless of whether they use the constant-shear zonal velocity or the Jovian zonal velocity shown in figure~\ref{fig:vzonal}.
 
 To create the 2-dimensional velocities needed for the stacked vortex at the collocation points $z \ne 0$, rather than use observations (which do not exist) or using other 2-dimensional solutions to \eqref{eq:a} and~\eqref{eq:b} (for which we have no guidance on what solutions to compute), we create new 2-dimensional velocities and vertical vorticities at the other collocation points by {\it projecting} the reference vertical vorticity $\omega_{z,vortex,0}(x,y) \equiv  \omega_{z,vortex}(x,y, z=0)$ to other heights $z$. (The 2-dimensional velocity at any $z$ is uniquely determined by $\omega_{z,vortex}(x,y, z)$ because $\nabla_{\perp} \cdot \bfv_{vortex} =0$ at all $z$.) For example, we could create a columnar stacked vortex aligned with the $z$ axis by defining $\omega_{z,vortex}(x,y,z)=\omega_{z,vortex,0}(x,y)$  for all $z$. Of course, the GRS is not an infinite column, and we require that it has a well-defined top and bottom. One way to do that is by  defining a family of vortices aligned with $z$ axis (that we call the $\mathrm{CA}\!-\!\mathrm{Rad}$ family):

\begin{eqnarray}
\mathrm{CA}\!-\!\mathrm{Rad} \,\, \mathrm{Family} \;\;\;\;\; \omega_{z,vortex}(x,y,z)&=&\omega_{z,vortex,0}(x,y)\, \frac{F(z)}{F(0)},\label{eq:init_method_CAr}
\end{eqnarray}
with
\begin{eqnarray}
   F(z)=\begin{cases}
   e^{-[(z-z_0)/H_{top}]^2} & \qquad (z>{z_0})\\
   e^{-[(z-z_0)/H_{bot}]^2} & \qquad (z<{z_0})
   \end{cases} \label{eq:fproject}
\end{eqnarray}
that projects $\omega_{z,vortex,0}(x,y)$ along the $z$ axis.  
Here, $F(z)$ is the projection function, which in this case determines how the magnitude of the vorticity changes as a function of $z$. The choice of this functional form of $F(z)$ is arbitrary, but note that an assumed Gaussian vertical dependence of the pressure, vertical vorticity, or angular velocity of the ocean and planetary vortices is quite common (c.f.,  \citet{yim2016stability, morales2013jupiter, mahdinia2017stability}).  We do not expect the $\omega_{z,vortex}$ of the final quasi-equilibrium vortex to have an exact Gaussian vertical dependence. However, by carefully choosing $F(z)$ and the other properties of the initial vortex, we are trying to ``nudge'' the final solution of our initial-value code to a hoped-for attracting basin that looks like a Jovian vortex. We use $F(z)$ and not $F(z_c)$ in all projections. Because  $z_c= \, z \, \sin \, \theta_0$, the two choices are equivalent when $H_{top}$ and $H_{bot}$ are re-scaled. We have labelled the family of initial conditions represented by equation~(\ref{eq:init_method_CAr}) as $\mathrm{CA}\!-\!\mathrm{Rad}$ because it projects $\omega_{z,vortex}$ along the planet's spherical {\it radial} axis $z$ (as in figure~\ref{fig:wz_stacked_rad}), while preserving the horizontal {\it area} of the vortex and decreasing the magnitude of $\omega_{z,vortex}(x,y)$ as $|z - z_0|$ increases.

To construct an initial condition stacked as in 
figure~\ref{fig:wz_stacked_rot} to create a family of vortices in the $\mathrm{CA}\!-\!\mathrm{Rot}$ family, then we set 
\begin{equation}
   \mathrm{CA}\!-\!\mathrm{Rot} \,\, \mathrm{Family} \;\;\;\;\; \omega_{z,vortex}(x,y,z)=\omega_{z,vortex,0}(x, y-z \, \cot \, \theta_0) \, \frac{F(z)}{F(0)},\label{eq:init_method_CA}
\end{equation}
where we have used the fact that equations~(\ref{eq:obey1}) and~(\ref{eq:obey2}) define the locus of point along a line parallel to the planet's spin axis $z_c$. 

Rather than requiring that the tops and bottoms of vortices be defined as the locations where their $\omega_{z,vortex}$ go to zero, we can define them by having their horizontal areas go to zero. To do that, we map the x-y plane at the cloud-top level to other depths via a hybrid coordinate $(x',y')$. For example,

\begin{eqnarray}
\mathrm{CV}\!-\!\mathrm{Rad} \,\, \mathrm{Family} \;\;\;\;\; \omega_{z,vortex}(x,y,z)&=&\omega_{z,vortex,0}(x',\, y')\label{eq:init_method_CVr}\\
x'&=&\frac{F(0)}{F(z)}x\\
y'&=&\frac{F(0)}{F(z)}y
\end{eqnarray}

produces a family of vortices, $\mathrm{CV}\!-\!\mathrm{Rad}$, oriented as in figure~\ref{fig:wz_stacked_rad},
where the magnitude of $\omega_{z,vortex}(x,y,z)$ is constant along $z$ but the horizontal area of the vortex decreases in a self-similar way from $z_0$ to the tops and bottoms of the vortices.
A similar family of initial vortices, $\mathrm{CV}\!-\!\mathrm{Rot}$, but oriented as in  figure~\ref{fig:wz_stacked_rot} can be created with

\begin{eqnarray}
  \mathrm{CV}\!-\!\mathrm{Rot} \,\, \mathrm{Family} \;\;\;\;\; \omega_{z,vortex}(x,y,z)&=&\omega_{z,vortex,0} \, (x',\, y')\label{eq:init_method_CV}\\
  x'&=&\frac{F(0)}{F(z)}x\\
  y'&=&\frac{F(0)}{F(z)}(y-z \, \cot \, \theta_0)
\end{eqnarray}

If we had used Moore-Saffman vortices, rather the modified velocity in equations~(\ref{eq:begin}) --~(\ref{eq:end}), then the 2-dimensional vortices at every height $z$ within the  $\mathrm{CV}\!-\!\mathrm{Rad}$  and $\mathrm{CV}\!-\!\mathrm{Rot}$ families would be exact solutions to equations~(\ref{eq:a}) --~(\ref{eq:c}) with $\beta=0$, rather than just good approximations, so we might expect that the initial vortices in these two families are close to the equilibrium solutions of the full equations. 

However, the vortices in the $\mathrm{CA}\!-\!\mathrm{Rad}$  and $\mathrm{CA}\!-\!\mathrm{Rot}$ families are far from equilibrium and not approximate solutions to equations~(\ref{eq:a}) --~(\ref{eq:b}). 
The shear $\sigma$ of the zonal flow at the tops and bottoms of these vortices is the same as it is at $z=0$, but the $\omega_{z,vortex}$ at the tops and bottoms is much smaller than it is at $z=0$. Thus  we expect that as these vortices evolve in time, their tops and bottoms become  stretched and elongated in the $x$ direction and possible destroyed. Our motivation for examining initial vortices of the $\mathrm{CA}\!-\!\mathrm{Rot}$ family in the next section is that \citet{galanti2019determining} and \citet{parisi2021depth} used them, or vortices similar to them,  as GRS models (they did not solve the governing equations of motion) to interpret {\it Juno} data.

\begin{figure}
\begin{subfigure}{0.45\textwidth}
\includegraphics[height=45mm]{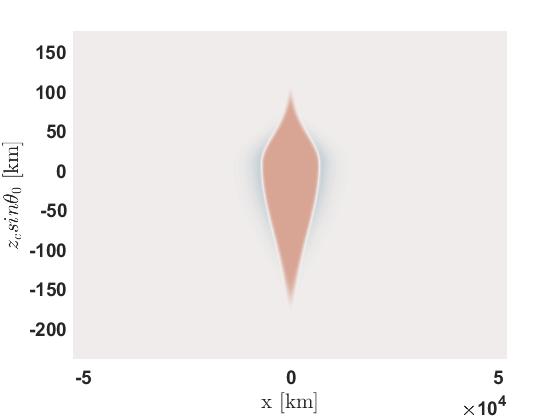}
    \caption{$\mathrm{CV}\!-\!\mathrm{Rot}$ vortex}
    \label{fig:wz_example_method_CV}
\end{subfigure}
\begin{subfigure}{0.53\textwidth}
\includegraphics[height=45mm]{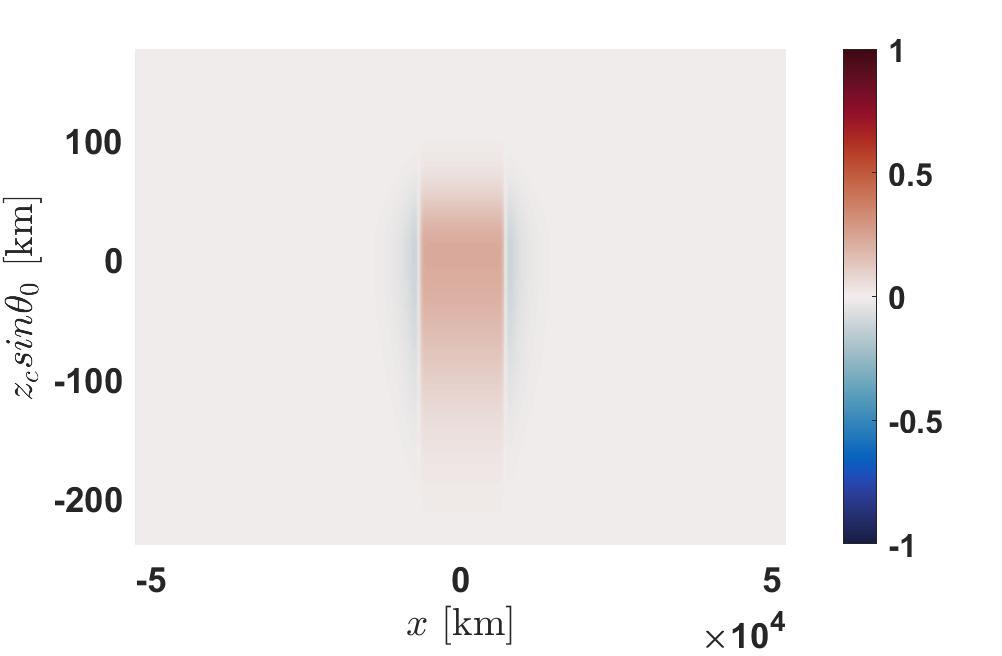}
    \caption{$\mathrm{CA}\!-\!\mathrm{Rot}$ vortex}
    \label{fig:wz_example_method_CA}
\end{subfigure}
\caption{Examples of the $\omega_{z,vortex}(x,y=0,z)/(f\sin\theta_0)$ of two initial vortices in the  $x$-$z$ plane at $y=0$. The horizontal axis is $x$, and the vertical is $z=z_c \, \sin \, \theta_0$.  Both vortices have $H_{top}=\,50$~km; $H_{bot} = \,120$~km; $\omega_0 =3.0\times 10^{-5}$s$^{-1}$; $\epsilon = 1.5$; $R_v =\,4,600$~km, and $L_r =\,2,000$~\,km.  (a) a constant-vorticity, area-varying vortex; (b) a constant-area, vorticity-varying vortex. Note that the planetary vortices are pancake vortices. For example, our vertical domain size is $\sim300$~km, while the horizontal domain size is $\sim100,000$~km. The figures are stretched along the vertical direction for graphical purposes. The vertical stretching applies to all figures of vertical slices presented.
\label{fig:wz_example_CAvsCV}}
\end{figure}

Examples of $\omega_{z,vortex}(x, y=0, z)$ for vortices from the 
$\mathrm{CV}\!-\!\mathrm{Rot}$  and $\mathrm{CA}\!-\!\mathrm{Rot}$ families are shown in Figure \ref{fig:wz_example_CAvsCV}. 

For all four families of vortices, $z=z_0$ is the height of the $x$-$y$ plane where the initial anticyclones have their maximum values of $\tilde{P}_{vortex}/\bar{\rho}$ and $|\bfv_{vortex}|$. It is also the $x$-$y$ plane where the initial $\mathrm{CA}$ vortices have their maximum values of $|\tilde{\omega}_z|$, and the $x$-$y$ plane where the initial $\mathrm{CV}$ vortices have their maximum values of area. We define the {\it mid-plane} $z_{mid}(t)$ of the vortex as the height $z$ where the anticyclone has its maximum $\tilde{P}_{vortex}/\bar{\rho}$ and note that $z_{mid}(t)$ is a function of time with $z_{mid}(t=0) = z_0$. Typically, $x$-$y$ plane at $z_{mid}(t)$ is where a vortex in the $\mathrm{CA}$ family will have its biggest vertical vorticity and where a vortex in the $\mathrm{CV}$ family will have its biggest horizontal area, so it is useful to track the flow fields in the mid-plane.

\section{Numerical Results with \boldmath{$\beta \equiv 0$}} \label{sec:numerical}

We examined several initial anticyclonic vortices from each of the four vortex families discussed in $\S$~\ref{sec:IC_method} with parameter values for $F(z)$ (in equation~(\ref{eq:fproject})) in the range: $120$~km~$\le H_{bot} \le 180$~km, $50$~km~$\le H_{top} \le 52.5$~km, and $0\le z_0 \le 15 $~km  (equivalently, between 500~mbar and 1~bar); and with parameter values for $\omega_{z,vortex,0}(x, y)$ (in equations~(\ref{eq:begin}) --~(\ref{eq:end})) in the range $-0.3 \le Ro \le -0.1$, or equivalently $1.2 \le \epsilon \le 1.5$. We chose $R_v = 4,600$~km and $L_r = 2000$~km. Here, the initial Rossby number is defined as $Ro \equiv \omega_0/(2f \, \sin \, \theta_0)$. Note that $\omega_0$ is slaved to the value of $\epsilon$ by equation~(\ref{eq:ms}) and the value of $\sigma$. In this paper we use $\sigma = -9.1\cross 10^{-6}\,\mathrm{s}^{-1}$ when calculating $\omega_0$ for all of the simulations, regardless of whether they use the constant-shear zonal velocity or the Jovian zonal velocity shown in figure~\ref{fig:vzonal}. We set $\theta_0 = 23^\circ$~S and $f = 3.5\times 10^{-4}\,\mathrm{s}^{-1}$. Throughout the paper, we report time in the unit of Earth days. Note that the figures of vertical slices are stretched vertically for graphical purposes. The simulated vortices are pancake vortices with very small vertical depths.

\subsection{Orientation: aligned with the $z$ or $z_c$ axis?}\label{sec:numerical_orientation}

\begin{figure}
    \centering
    \includegraphics[trim={2cm 0 0 0},clip,width=140mm]{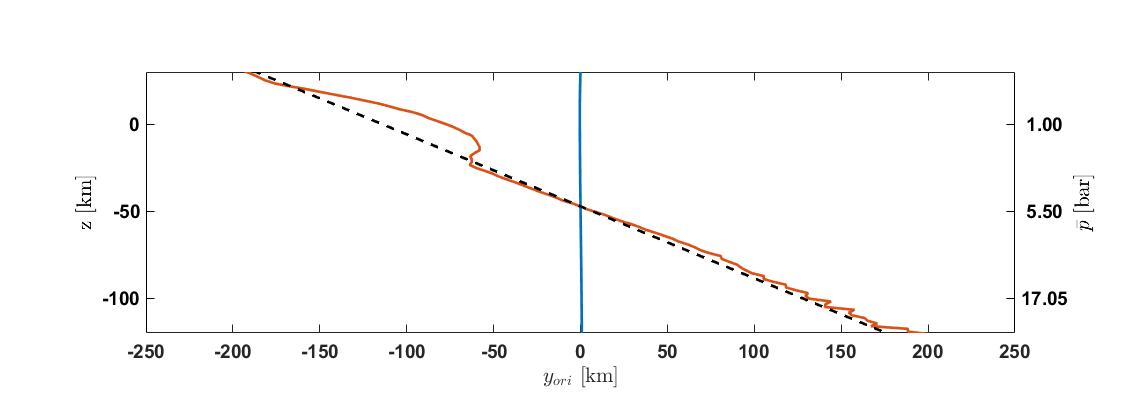}
    \caption{The change in the orientation of the central axis, $y_{ori}(z)$ of a vortex from initial condition to final state. The black dashed line has slope $\tan \, \theta_0$ and is parallel to the planet's spin, or $z_c$ axis. The blue vertical line shows the orientation of the initial vortex: $y_{ori} \equiv 0$ for all $z$, which is parallel to the direction of gravity or $z$ axis. The continuous red curve shows $z$ as a function of $y_{ori}$ for the final vortex. Its central axis is aligned with the spin axis of the  planet for almost all $z$, with the exception of the mid-plane near $z = -20$~km.  
    The initial vortex is part of the $\mathrm{CV}\!-\!\mathrm{Rad}$ family, is in near-equilibrium, and constructed using equation~(\ref{eq:init_method_CVr}) with $H_{bot} = 120$~km, $H_{top} = 50$~km, and $z_0 = 8.2$~km. For constructing $\omega_{z, vortex, 0}(x,y)$, we chose $Ro = -0.1$, or equivalently $\epsilon = 1.5$.}
    \label{fig:orientation_change}
\end{figure}

The planetary vortices of interest in this study have low Rossby number ($Ro\lesssim 0.3$), and all the vortices that we examined have their final axes approximately aligned with the planet's spin axis ($z_c$ axis), even if they were initially aligned with the local gravity ($z$ axis). This is consistent with our previous findings \citep{zhang2022hydrodynamic}. For example, figure~\ref{fig:orientation_change} shows how the central axis of a vortex initially in the $\mathrm{CV}\!-\!\mathrm{Rad}$
family changes from the beginning to the  end of a simulation. For all heights $z$, we define the $y$ location of the the vortex center as 
\begin{eqnarray}
y_{ori}(z)=\frac{\iint[\tilde{P}_{vortex}(x,y,z)/\bar{\rho}(z)] \,y\,dxdy}{\iint[\tilde{P}_{vortex}(x,y,z)/\bar{\rho}(z)] \, dxdy}
\label{eq:orientation_axis_def}
\end{eqnarray}

We use $\tilde{P}_{vortex}/\bar{\rho}$ as the weighting factor because it is almost never negative and is a good proxy for the stream function of the horizontal velocity. If $dz/dy_{ori} = \pm \infty$, then the central axis is aligned with $z$ axis, and (using equation~(\ref{eq:obey2})) if  $dz/dy_{ori} = \tan \, \theta_0$, it is aligned with the $z_c$ axis. Initially, the vortex in figure~\ref{fig:orientation_change} is aligned with the $z$ axis, but by the end of the simulation for most values of $z$, the slope $dz/dy_{ori} \simeq \tan \, \theta_0$, and the vortex is aligned with the $z_c$ axis. However it must be noted that because the vertical thickness of the vortex is so much smaller than its diameter ($\sim 0.01$), the change in the location of the central axis during the simulation is extremely small. 

The red curve defining $y_{ori}(z)$ in figure~\ref{fig:orientation_change} does not align perfectly with the $z_c$ axis. A small portion of $y_{ori}(z)$ (approximately, $-20$~km $< z<0$~km) is unaligned. 
Our simulations show that whether the central axis of a vortex aligns with the direction of the local gravity ($z$ axis) or with the $z_c$ axis depends on more than the value of the local Rossby number. However, we postpone that discussion for a future paper. All of our numerical simulations with initial vortices in the  $\mathrm{CV}\!-\!\mathrm{Rad}$ and $\mathrm{CA}\!-\!\mathrm{Rad}$ families ended with their central axes approximately aligned with the $z_c$ axis. For this reason, we confine the remainder of this section to the evolution of vortices that are initially in the  $\mathrm{CV}\!-\!\mathrm{Rot}$ and $\mathrm{CA}\!-\!\mathrm{Rot}$ families.

\subsection{$\mathrm{CV}\!-\!\mathrm{Rot}$} \label{sec:numerical_CV}
Using equation~(\ref{eq:init_method_CV}), the maximum vertical vorticity $\omega_{z,vortex}$ of the initial vortices is the same at each height, but the area of the vortex changes. We examined several initial $\mathrm{CV}\!-\!\mathrm{Rot}$ vortices, and in all cases, the initial vortex converges to a large, quasi-steady vortex that drifts longitudinally. The vertical size of the vortex remains similar to its initial condition. An initially deeper $\mathrm{CV}\!-\!\mathrm{Rot}$ vortex evolves to a deeper final state unless there is convectional instability locally (See section \ref{sec:numerics_convection} for details).\par 

To illustrate how this family of vortices evolves, $\S$~\ref{sec:caseCV1} shows an example of an initial vortex embedded in a constant-shear zonal velocity (Case $CV1$).  In $\S$~\ref{sec:caseCV2}, we examine the same initial vortex embedded in the observed Jovian zonal velocity (Case $CV2$).  Both cases have $H_{bot} = 120$~km, $H_{top} =  50$~km,$\epsilon = 1.5$, (equivalently,  $Ro = -0.1$), and $z_0 = 8.2$~km (equivalently, $700$~mbar; note that  $z=0$, the top of the visible clouds,  corresponds to  $\bar{P} \simeq 1$~bar). One reason that we choose this value $z_0$ was so that there are no initial, local convective instabilities (see $\S$~\ref{sec:numerics_convection}).

\subsubsection{Vortex Embedded within a Zonal Velocity with Constant Shear (Case $CV1$)\label{sec:caseCV1}}
\begin{figure}
\includegraphics[width=135mm]{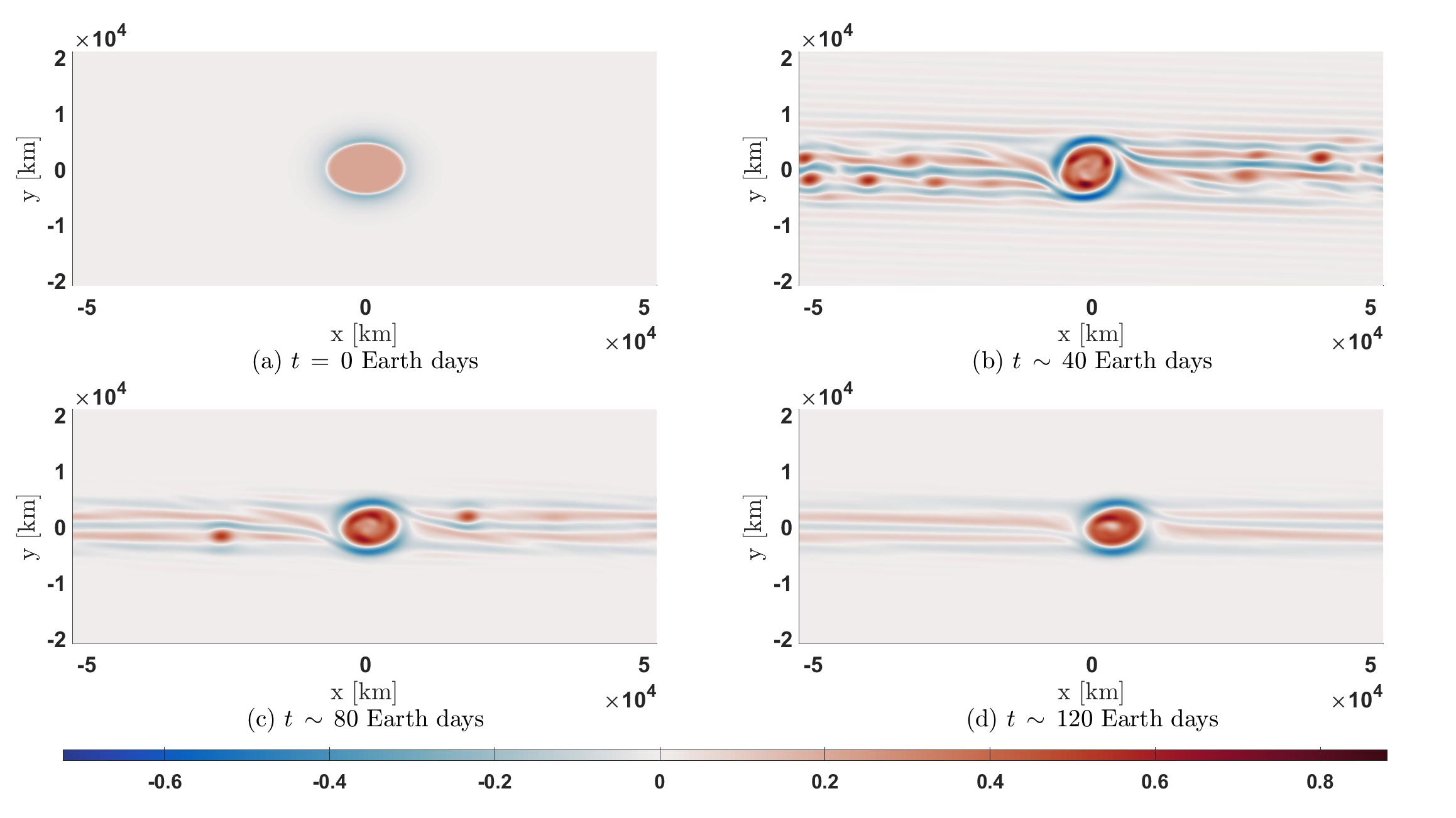}
\caption{ $\omega_{z,vortex}(x,\,y)/|f\sin\theta_0|$ at the mid-plane of the vortex. (Case $CV1$). 
At all times, the vortex mid-plane is defined as the height $z$ where the anticyclone has its maximum $\tilde{P}_{vortex}/\bar{\rho}$. 
The time is in units of  Earth days The color bar shows the value of $\omega_{z,vortex}(x,\,y)/|f\sin\theta_0|$. The turn-around time of this vortex is $\sim 4$~ days; that of the GRS is $\sim$~6 days \citep{marcus1993jupiter}. In the first 40 days, the vortex remains mostly intact, but casts off many filaments that roll-up into small vortice that are similar to those in the $\mathrm{CV}\!-\!\mathrm{Rot}$ family. After 120 days, all of the small vortices have merged with the large vortex. The center of the vortex is prominently hollow with a local minimum of $\omega_{z,vortex}$ at its center. The vortex remains shielded, but most of the negative-$\omega_{z,vortex}$ has moved to the vortex's northern and southern edges. We illustrate this case because it is qualitatively similar to all the simulations we carried out with $\mathrm{CV}\!-\!\mathrm{Rot}$ vortices embedded in constant zonal shear. There is little qualitative change of the vortex after 120 days. \label{fig:wz_xOy_caseCV1}}
\end{figure}

\begin{figure}
\includegraphics[width=135mm]{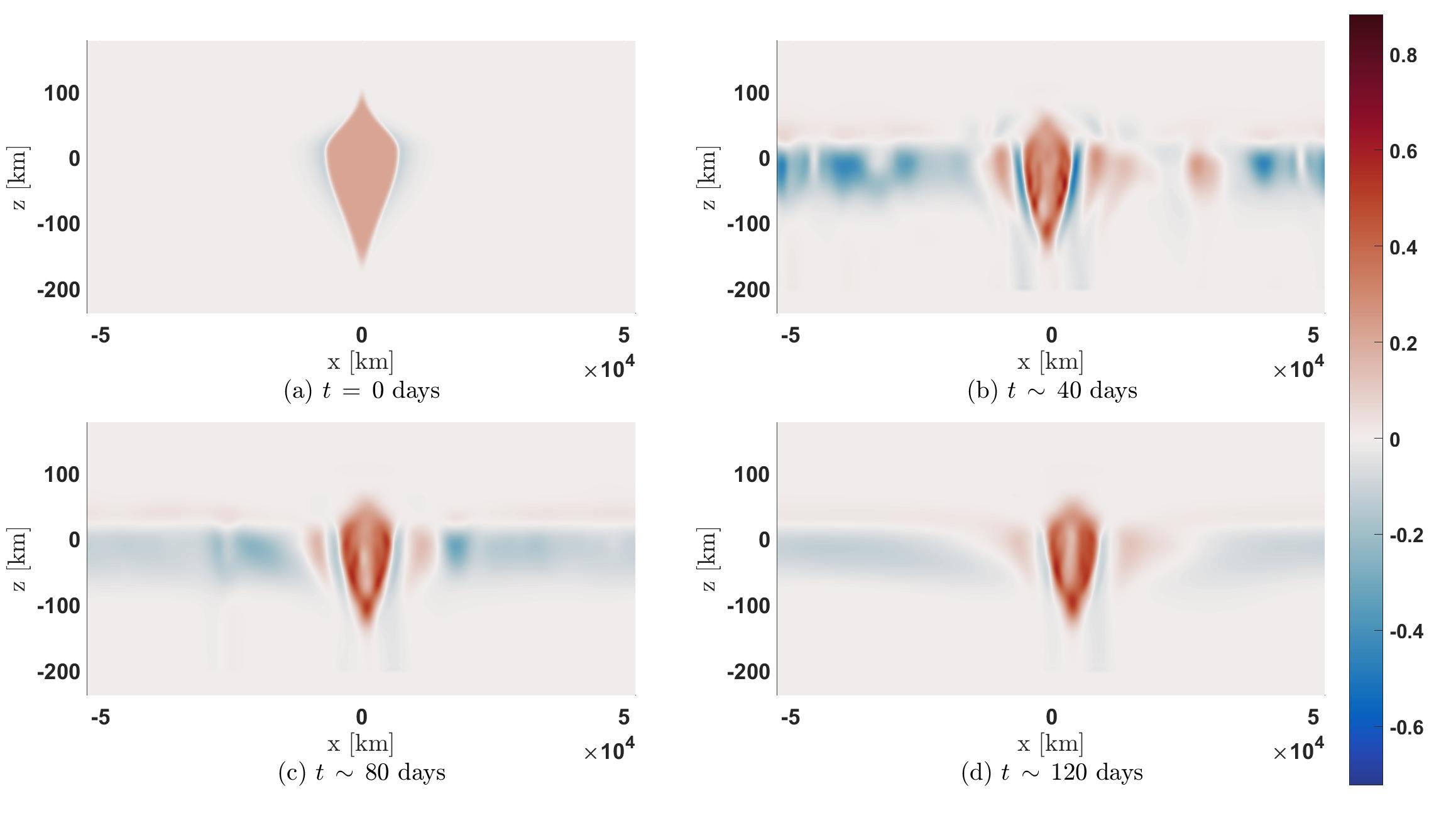}
\caption{$\omega_{z,vortex}(x,\,y=0,\,z)/|f\sin\theta_0|$ in the x-z plane of the vortex (Case $CV1$). The color bar is plotted as in figure~\ref{fig:wz_xOy_caseCV1}. In this plane at day 40, the shed filaments have mostly negative values of $\omega_z$, and the hollowness along the central axis of the vortex has developed and extends from near the top of the vortex to near the bottom of the vortex. At day 120, some of the negative $\omega_z$ in the initial shield from other $x$-$z$ planes has become concentrated in this plane at $y=0$ and fills much of the east-west domain.\label{fig:wz_xOz_caseCV1}}
\end{figure}

Figure~\ref{fig:wz_xOy_caseCV1} shows $\omega_{z,vortex}/|f\sin\theta_0|$
in the $x$-$y$ plane at the mid-plane at four different times. Because the initial condition is not in exact equilibrium, parts of it are torn apart by the zonal flow and roll up into small vortices. However, most of the initial vortex resists the zonal shear and remains intact. By the end of the simulation, all of the small vortices have merged back into the main vortex or becomes sheared out in the $x$ direction modifying the zonal flow, and there is only one large longitudinally-drifting vortex. Figure \ref{fig:wz_xOz_caseCV1} shows $\omega_{z,vortex}/|f\sin\theta_0|$ in the $x$-$z$ plane at $y=0$, which contains the center axis of the vortex. Throughout the simulation, the heights of the top and bottom of the vortex remain approximately constant. The hollowness, or local minimum of $\omega_{z,vortex}$, at the central rotation axis of the vortex is visible at most values of $z$.

The truly surprising result is that the vortex becomes 
hollow. The initial $\omega_{z,vortex}$ in figures~\ref{fig:wz_xOy_caseCV1}a and~\ref{fig:wz_xOz_caseCV1}a is non-hollow, but in the center and along the central spin axis of the vortices in figures~\ref{fig:wz_xOy_caseCV1}bcd and ~\ref{fig:wz_xOz_caseCV1}bcd, $\omega_{z,vortex}$ is smaller than it is in the outer parts of the vortices. In our previous 2-dimensional, quasigeostrophic studies of vortex dynamics, we never found an initially non-hollow vortex spontaneously becoming hollow. Moreover, the only way we found to prevent an initially hollow vortex from becoming non-hollow was by adding an artificial ``bottom topography'' to the governing quasigeostrophic equation of motion \citep{youssef2000dynamics,shetty2006modeling}. However, \citet{barranco2005three} found stable hollow vortices in 3D protoplanetary disk simulations, which indicates that the hollow vortex structure is indeed a three-dimensional effect.


\begin{figure}
\includegraphics[width=135mm]{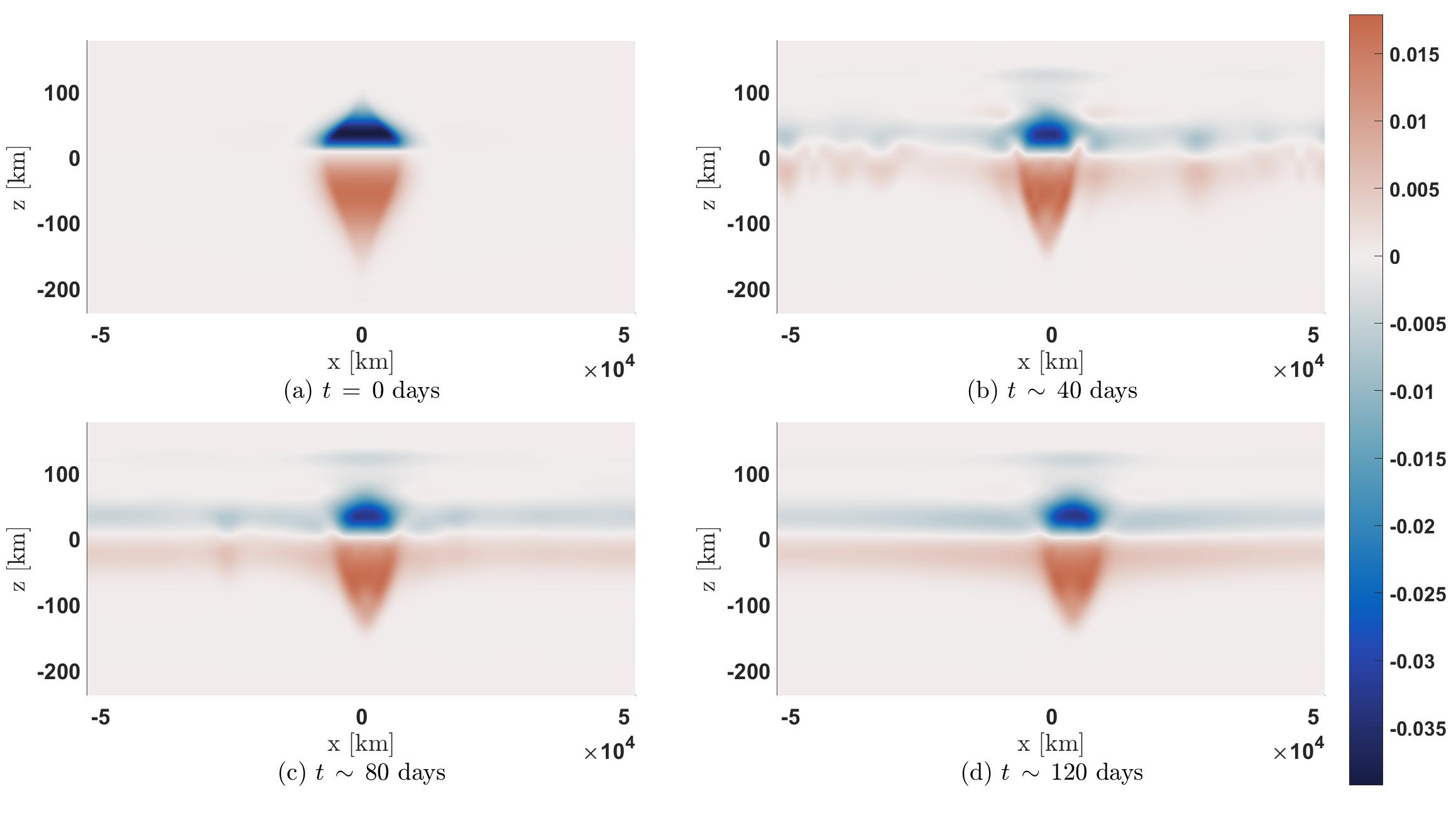}
\caption{$\tilde{\Theta}_{vortex}(x,\,y=0,\,z)/\bar{\Theta}$ (with values as shown in the color bar) in the $x$-$z$ plane passing through the central axis of the vortex (Case $CV1$). The mid-plane of the vortex lies in the horizontal band of white where $\tilde{\Theta}_{vortex}(x,\,y,\,z) \simeq 0$. The high-$\tilde{\Theta}_{vortex}$ annular ring at the edge of the vortex can be seen in panels (b) --~(d) by noting that the white horizontal band has high ``shoulders'' at the two edges of the vortex. The shoulders indicate that they contain higher $\tilde{\Theta}_{vortex}/\bar{\Theta}$ than the surrounding fluid at the same $z$. The shoulders are not present in panel (a) at $t=0$. The top of the vortex in panel (d) is at $z=60$~km, which corresponds to an atmospheric pressure of $35$~mbar. The tops of the large Jovian vortices are thought to be at a height above 140~mbar \citep{fletcher2010thermal} or heights above $z=37$~km in this figure.}  \label{fig:ptemp_xOz_caseCV1}
\end{figure}

Figure \ref{fig:ptemp_xOz_caseCV1} shows  $\tilde{\Theta}_{vortex}/\bar{\Theta}$ in the $x$-$z$ plane passing through the center of the vortex. The cool-top and warm-bottom of the vortex occur because the major force balance in the $z$ component of the momentum equation is hydrostatic balance

\begin{eqnarray}
\pdv{}{z}\,\left(\frac{\tilde{P}_{vortex}}{\bar{\rho}}\right)\sim \frac{\tilde{\Theta}_{vortex}}{\bar{\Theta}}g. \label{eq:near_hydrostatic_eq}
\end{eqnarray}

An anticyclone has a high pressure center, so $\tilde{\Theta}_{vortex}<0$ at heights above the mid-plane, and $\tilde{\Theta}_{vortex}>0$ below the mid-plane to maintain the balance. The cooler temperature of the top of the  GRS and other Jovian anticyclones has been observed many times \citep{flasar1981thermal,fletcher2010thermal,fletcher2016mid}. 

\begin{figure}
\includegraphics[width=135mm]{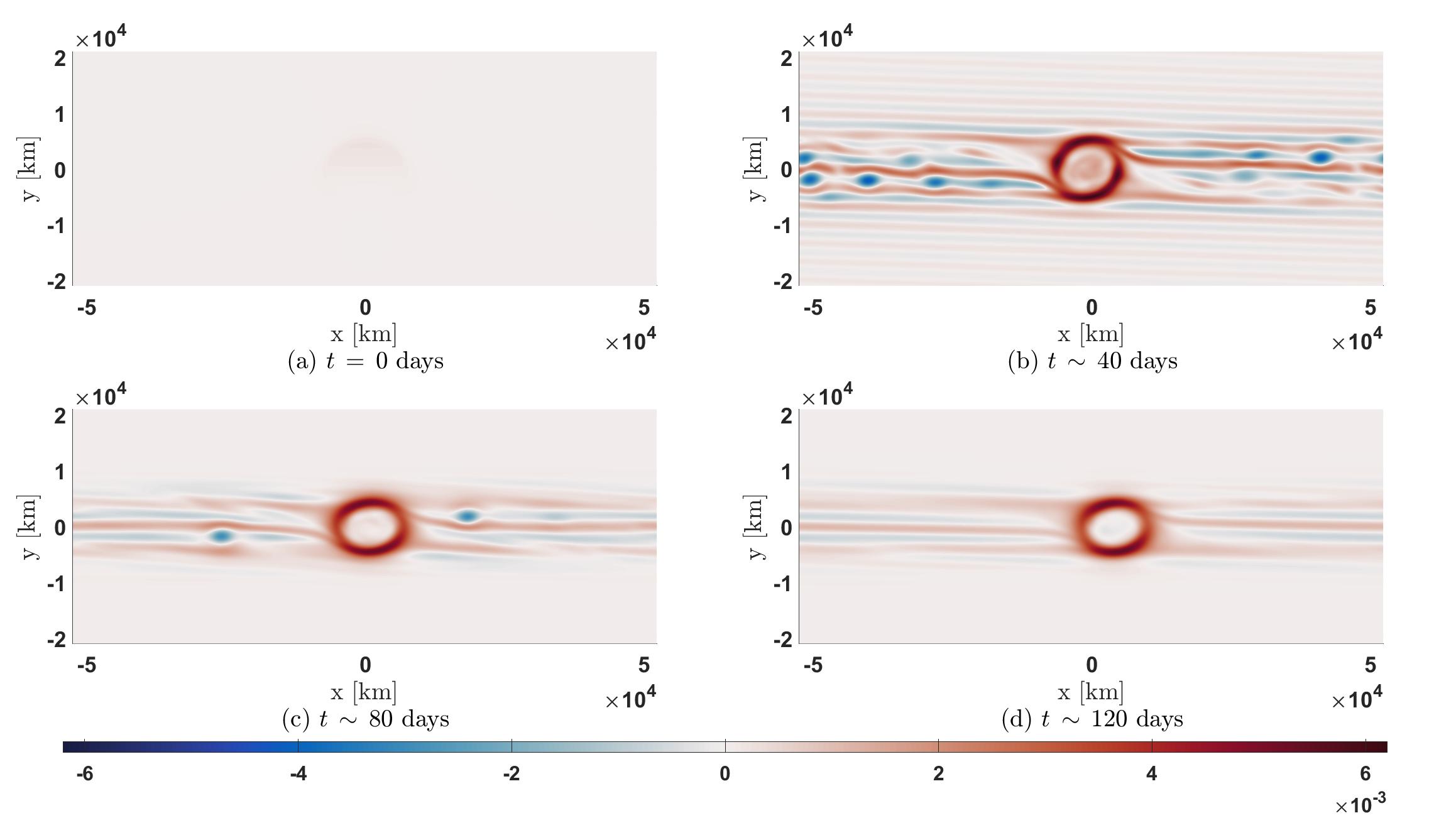}
\caption{$\tilde{\Theta}_{vortex}(x,\,y)/\bar{\Theta}$ on the mid-plane of the vortex (Case $CV1$). This plane corresponds the white horizontal band in the middle of the vortex in figure \ref{fig:ptemp_xOz_caseCV1}. The characteristic difference between the temperature in the high-$\tilde{\Theta}_{vortex}$ annular ring at the outer edge of the vortex and the ambient flow is  approximately 1\,K. The high--$\tilde{\Theta}_{vortex}$ annulus in panels (b) --~(d) is not present in panel (a) at $t=0$. The high-$\tilde{\Theta}_{vortex}$ annular ring at the edge of the vortex corresponds to the high ``shoulders'' in  figure~\ref{fig:ptemp_xOz_caseCV1}bcd.   \label{fig:ptemp_xOy_caseCV1}}
\end{figure}

Figure \ref{fig:ptemp_xOy_caseCV1} shows  $\tilde{\Theta}_{vortex}(x,\,y)/\bar{\Theta}$ in the mid-plane of the vortex, corresponding to the horizontal white region in the middle of figure~\ref{fig:ptemp_xOz_caseCV1}. 
It is not known whether the mid-planes of the Jovian vortices are at, below, or above the visible cloud-top level at $z=0$ (i.e., at $\sim$~1~bar). Initially, the mid-planes of our vortices are at $z_0$, and although the height of the mid-plane changes in time, they remain near $z_0$. Figure~\ref{fig:ptemp_xOy_caseCV1} shows a thin annulus of high  $\tilde{\Theta}_{vortex}/\bar{\Theta}$ at the outer edge of the vortex, approximately coincident with the shield of negative $\omega_z$. This is an unexpected result, and the figure shows that this annulus was not present in the initial conditions. 
Because there is no corresponding anomaly in $\tilde{P}_{vortex}/\bar{P}$ at the location of this annulus (and because $\tilde{T}_{vortex}/\bar{T}$ = $\tilde{\Theta}_{vortex}/\bar{\Theta}$ + $(R/C_p) \, \tilde{P}_{vortex}/\bar{P}$), there is a high-temperature annulus of $\tilde{T}_{vortex}/\bar{T}$ at the outer edge of the vortex where the characteristic temperature of the annulus is approximately 1~K warmer than the surrounding fluid. Because this annulus was not present initially and because $N^2 > 0$, the energy equation~(\ref{eq:anelastic_energy}) shows that the annulus of high $\tilde{\Theta}_{vortex}$ was created by a {\it downward} vertical 
velocity that formed in the annulus. That is, the high-temperature annulus was created by adiabatic heating. \citet{de2010persistent} observed $5$~$\mu$m bright annular rings around the GRS and other Jovian anticyclones and hypothesized (1) that the bright rings were created by down-welling velocities in the annuli that could not be directly observed; (2) that the adiabatic heating caused by the unseen down-welling locally dried out the atmosphere making the annular regions free of clouds; and (3) that the $5$~$\mu$m radiation from the underlying, high-temperature atmosphere, which would normally be blocked by the clouds, would be observed. Figure~\ref{fig:ptemp_xOy_caseCV1} supports these hypotheses. 

\begin{figure}
    \centering
    \includegraphics[trim={2.5cm 0 2.5cm 0},clip,width=135mm]{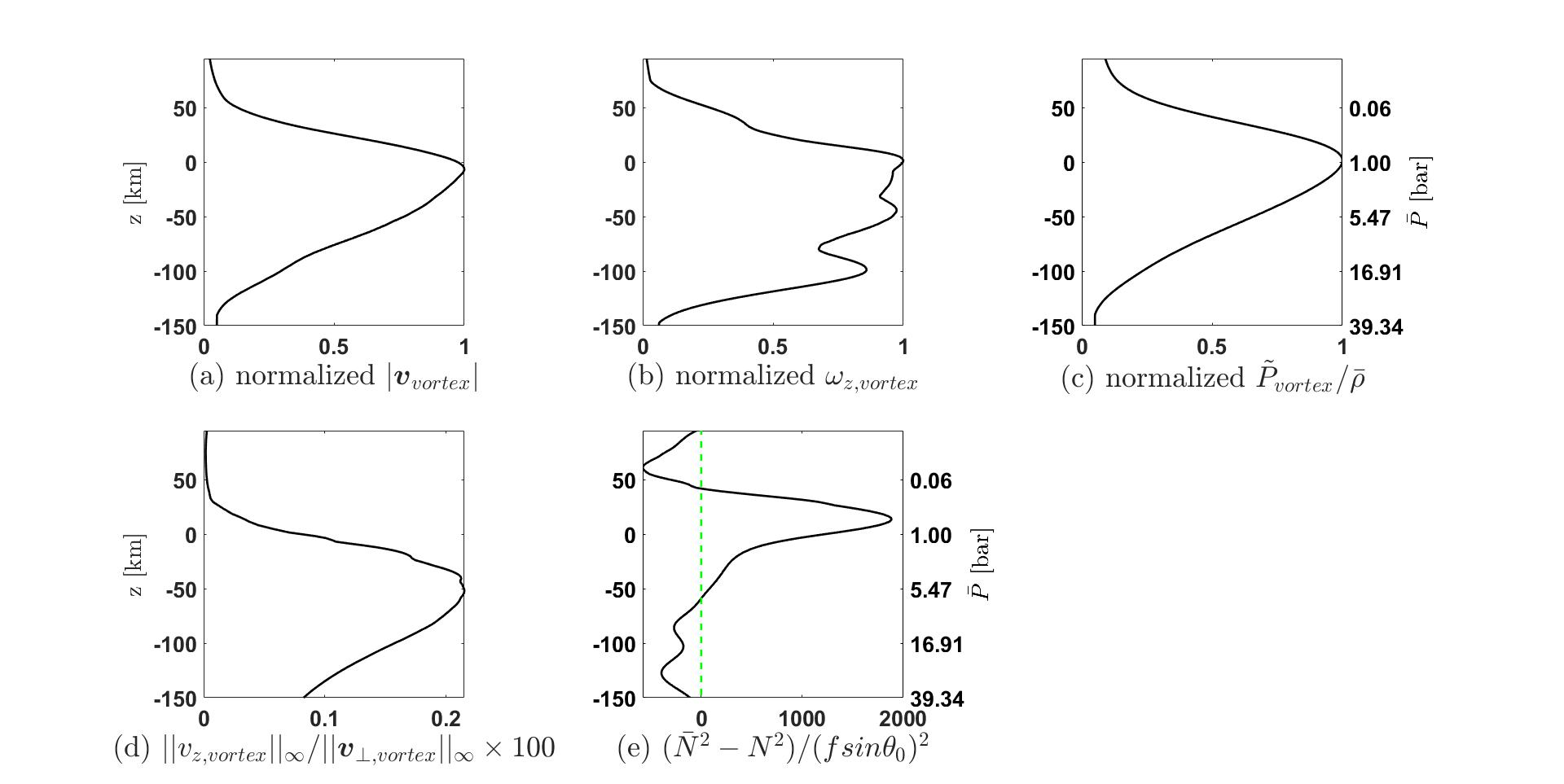}
    \caption{Vertical profiles of the vortex in case $CV1$. Panel a: maximum $|\bfv_{vortex}|$ at each x-y plane as a function of z. Panel b: maximum $\omega_{z,vortex}$ at each x-y plane as a function of z. Panel c: maximum $\tilde{P}_{vortex}/\bar{\rho}$ at each x-y plane as a function of z. Panel d: the maximum $|{v}_{z,vortex}|$ at each x-y plane as a function of z.  Panel e: $(\bar{N}^2-N^2)$
    along the z axis of the vortex. Panels (a) --~(c) are normalized by their maximum values. Panel (d) is normalized by the maximum $|\bfv_{vortex}|$ of the vortex. The green dashed line in panel (e) indicates where  $(\bar{N}^2-N^2)$. Panel (e) shows that near the mid-plane the vortex is de-stratified, making it more prone to convective instabilities; whereas at the vortex's top and bottom, the fluid is superstratified and more resistant to convection. Note that the mid-plane location is $z=0.58$~km ($975$~mbar) at day 120, slightly lower than the height of its initial location at $z_0=8.2$~km ($700$~mbar). See text for details.} 
    \label{fig:vertical_profiles_CV1}
\end{figure}

Figure~\ref{fig:vertical_profiles_CV1} shows the maximum values of 
$|\bfv_{vortex}|$, $\omega_{z,vortex}$, $\tilde{P}_{vortex}/\bar{\rho}$, and $v_{z,vortex}$ in each $x$-$y$ plane as functions of $z$ for the quasi-steady final vortex at day 120. The figure also shows $(\bar{N}^2 - N^2)$ as a function of $z$ along the central axis of the vortex.  Panels a and c show that the vertical profiles of $|\bfv_{vortex}|$ and $\tilde{P}_{vortex}/\bar{\rho}$ are similar, as would be expected from geostrophic balance in equation~(\ref{eq:anelastic_momentum}). 
The mid-plane of the final vortex at 120 days is not very different from its initial location at $z=0$. Panel~b shows that the the vertical vorticity does not have a sharp maximum at the vortex mid-plane, but is nearly constant from $z=0$ down to $z= -100$~km. This is evidence that the final vortex is a $\mathrm{CV}\!-\!\mathrm{Rot}$ vortex. Panel d shows that $v_z$ is $0.2\%$ of the full velocity at most in the three-dimensional simulation. Panel e shows $(\bar{N}^2 - N^2) = - (g/\bar{\Theta}) \, (\partial \tilde{\Theta}/\partial z)$
which is the stratification due to the net velocity (zonal flow + vortex) along the central axis of the vortex. The value is nondimensionalized by $(f\sin\theta_0)^2$. When this quantity is greater than zero, it means that the vortex has locally mixed the fluid in a way to de-stratify the vortex; the mixing has increased the potential temperature at the bottom of the vortex at the expense of the potential temperature at the top of the vortex. (In other words, mixing has increased the potential mass density at the top of the vortex at the expense of the mass density at the bottom of the vortex, making it more prone to local convective instabilities). In regions where $\tilde{N}^2_{vortex}/g < 0$, mixing has super-stratified the fluid, making it more stable to convection.


\subsubsection{Vortex Embedded in the Jovian Zonal Velocity (case CV2)\label{sec:caseCV2}}

\begin{figure}
\includegraphics[width=135mm]{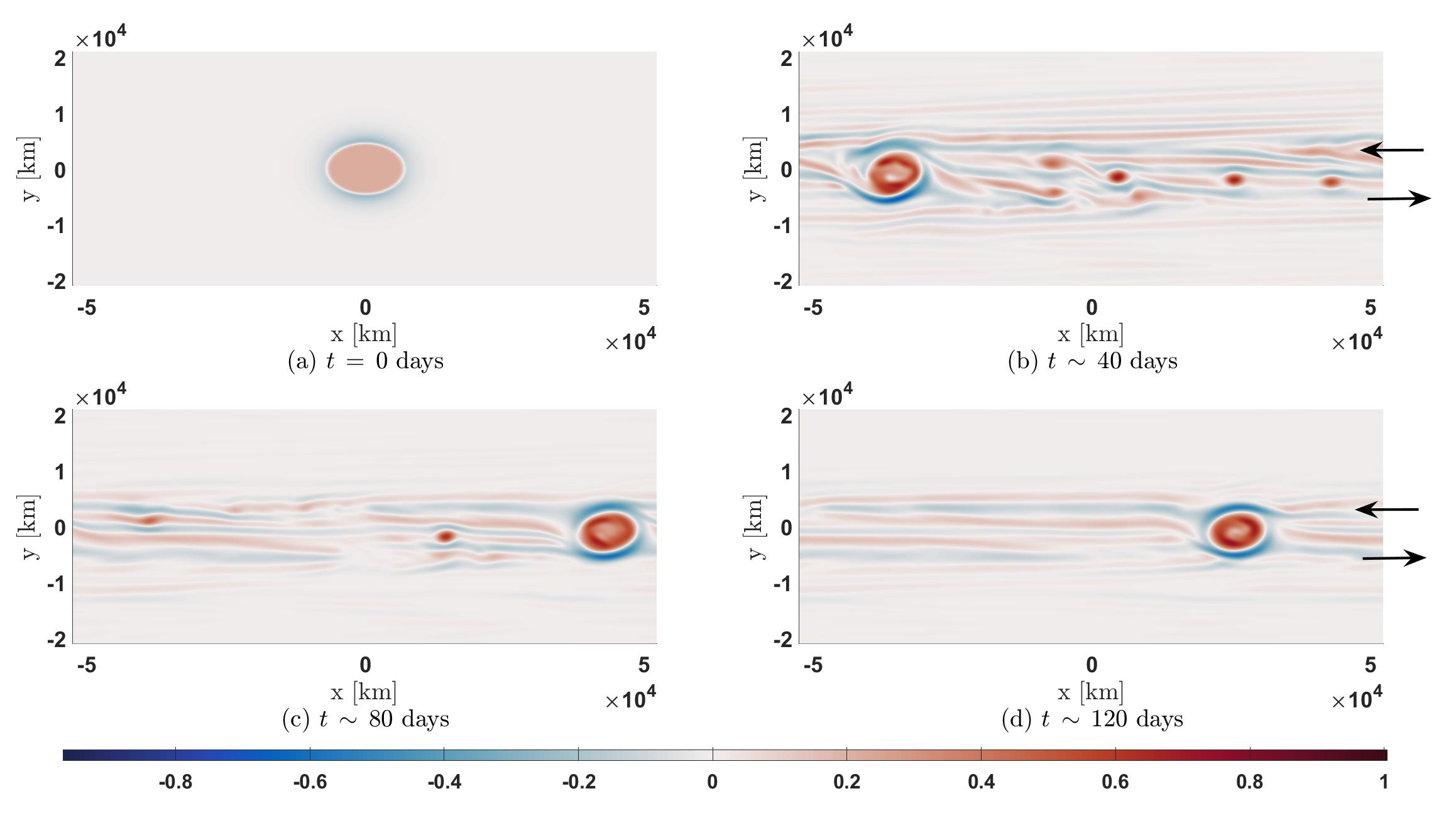}
\caption{$\omega_{z,vortex}(x,\,y)/|f\sin\theta_0|$ of the $CV2$ vortex plotted as we did for the $CV1$ vortex in figure~\ref{fig:wz_xOy_caseCV1}.  The two horizontal arrows at the right of the figure indicate the latitudes where the zonal flow in figure~\ref{fig:vzonal} has local maximum and minimum values. The arrows point in the direction of the local zonal flow. \label{fig:wz_xOy_caseCV2}}
\end{figure}

\begin{figure}
\includegraphics[width=135mm]{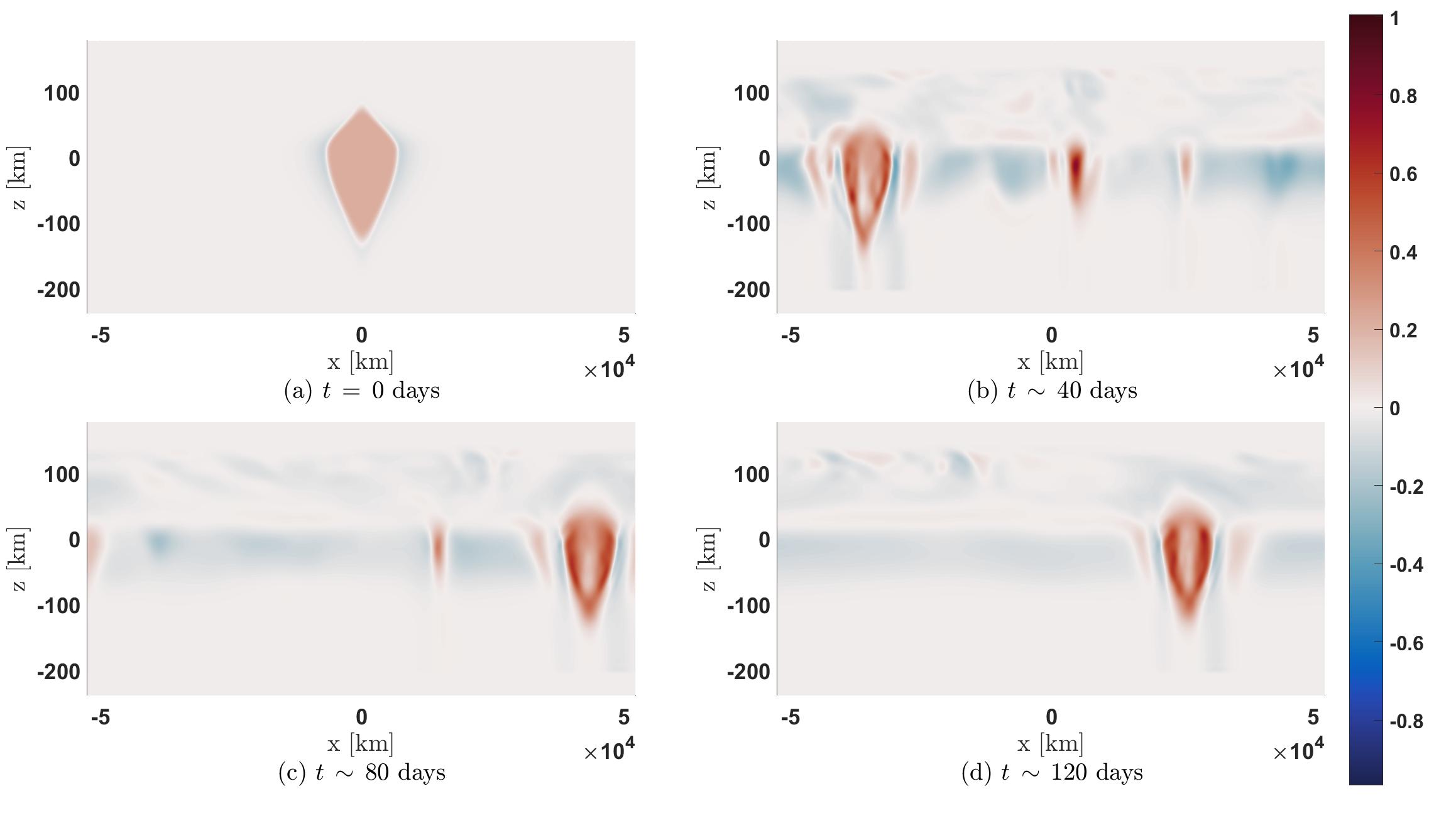}
\caption{$\omega_{z,vortex}(x,\,y=0,\,z)/|f\sin\theta_0|$ of the $CV2$ vortex plotted as we did for the $CV1$ vortex in figure~\ref{fig:wz_xOz_caseCV1}. \label{fig:wz_xOz_caseCV2}}
\end{figure}

\begin{figure}
\includegraphics[width=135mm]{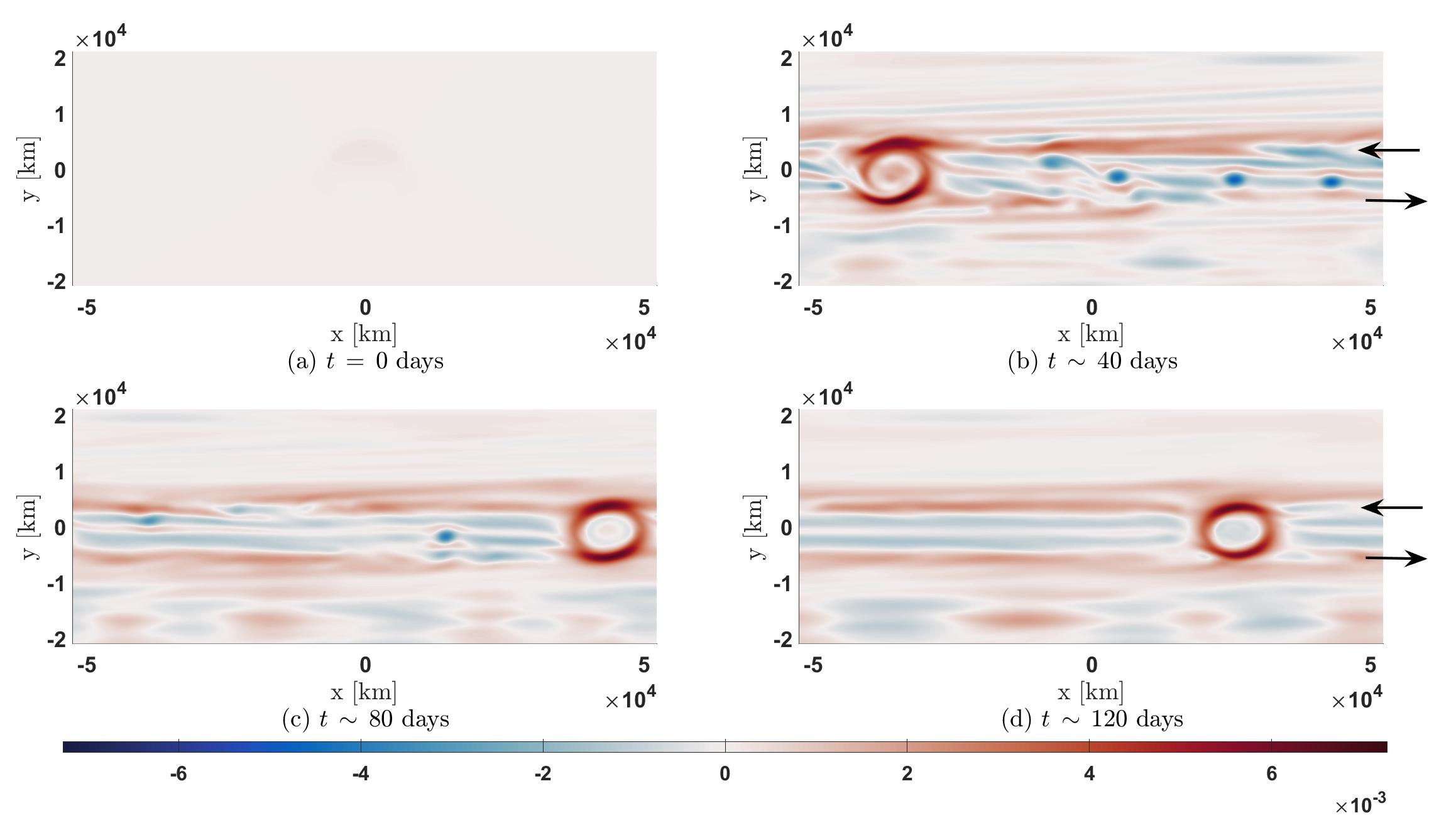}
\caption{$\tilde{\Theta}(x,\,y)/\bar{\Theta}$ of the $CV2$ vortex plotted as we did for the $CV1$ vortex in figure~\ref{fig:ptemp_xOz_caseCV1}. The black arrows indicates the location of the westward/eastward zonal flow's maximum location\label{fig:ptemp_xOy_caseCV2}}
\end{figure}

Note that case CV1 is conducted in a constant-shear zonal flow without the locations of maximum and minimum zonal velocity. On the other hand, \citet{shetty2007interaction} suggested that the latter is important to make hollow vortices in stable equilibrium in quasi-geostrophic system with bottom topography. We repeated the numerical calculation with the same initial vortex that we used in the $CV1$ case, but embedded in the Jovian zonal velocity (blue curve shown in  figure~\ref{fig:vzonal}), rather than with the straight red line (which has no local maximum or minimum velocities). The purpose of this simulation is to see if including the location of maximum and minimum zonal velocity can significantly affect the simulation result. Figures~\ref{fig:wz_xOy_caseCV2} --~\ref{fig:ptemp_xOy_caseCV2} shows the vertical vorticity and potential temperature of this new vortex labeled $CV2$. To our surprise, the two final vortices are qualitatively similar, other than the fact that the horizontal area of $\omega_z$ of the vortex is smaller than that of the $CV1$ vortex at day 120. They both are shielded and both developed warm, thin annular rings at their outer edges like the large Jovian anticyclones. In addition, they both developed hollow interiors like the GRS. Our result indicates that the location (or existence) of the maximum and minimum zonal velocity is not important for the vortex features we capture in both simulations (hollowness, warm ring, etc.)


\begin{figure}
\includegraphics[trim={2cm 0 2cm 0},clip,width=135mm]{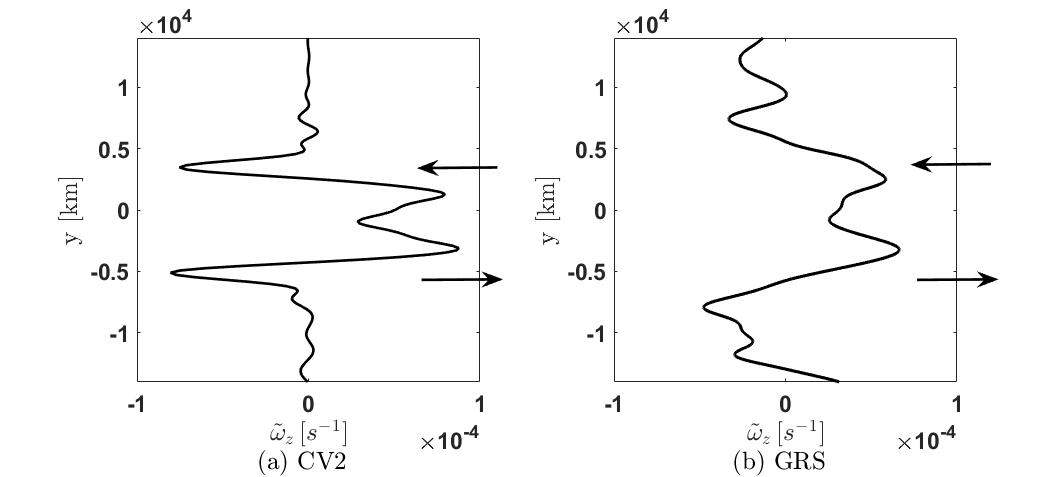}
\caption{Comparison of the $\omega_{z,vortex}$ of the $CV2$ vortex at the cloud-top plane and the observed GRS along the principle, minor, north-south or $y$-axis. Panel (a) is $CV2$. Panel (b) is the observed GRS profile from \citet{wong2021evolution}. The black arrows indicates the locations of the extrema of westward/eastward zonal velocity. 
The hollowness ratio (defined as the maximum value of $\omega_{z,vortex}$ in the ``shoulders'' divided by its local minimum value at the vortex center -- see text for details) of the GRS and $CV2$ are both $\sim$~2. The main difference between the GRS and $CV2$ is that the GRS is larger.  \label{fig:wz_vs_GRS_cases}}
\end{figure}

Although we did not intend to recover the velocity and temperature field of any specific Jovian vortex in this study, the velocities of the $CV1$ and $CV2$ quasi-steady vortices are qualitatively similar to the GRS because of their hollow interiors. 
This can be seen in figure~\ref{fig:wz_vs_GRS_cases}, which shows the $\omega_{z,vortex}$ of the GRS and of the $CV2$ vortex along their north-south minor principle axes as functions of $y$, at the cloud-top plane.   
The interiors of the vortices are the regions in  $y$  where 
$\omega_{z,vortex} >0$; the shields are the regions just outside interiors where $\omega_{z,vortex} <0$. The $\omega_{z,vortex}$ of a  non-hollow shielded vortex would monotonically decrease as a function of $|y|$ from its maximum value at the latitude of the vortex center at $y=0$ until it became negative in the shield, and then it would increase to zero at the outer edge of the shield.  Both the GRS and $CV2$ have pronounced local minima of $\omega_{z,vortex}$ at $y=0$ surrounded in the north and south by large ``shoulders'' of positive $\omega_{z,vortex}$. A measure of the hollowness of a vortex is the ratio of the largest value of $\omega_{z,vortex}$ in the shoulders divided by the local minimum value of $\omega_{z,vortex}$ at the vortex center. Both the GRS and $CV2$ have similar ratios of $\sim$~2. However, the GRS is larger than $CV2$, and its region of hollowness is also larger,

\subsection{$\mathrm{CA}\!-\!\mathrm{Rot}$} \label{sec:numerical_CA}
\begin{figure}
\includegraphics[trim={1cm 0 0 0},clip,width=135mm]{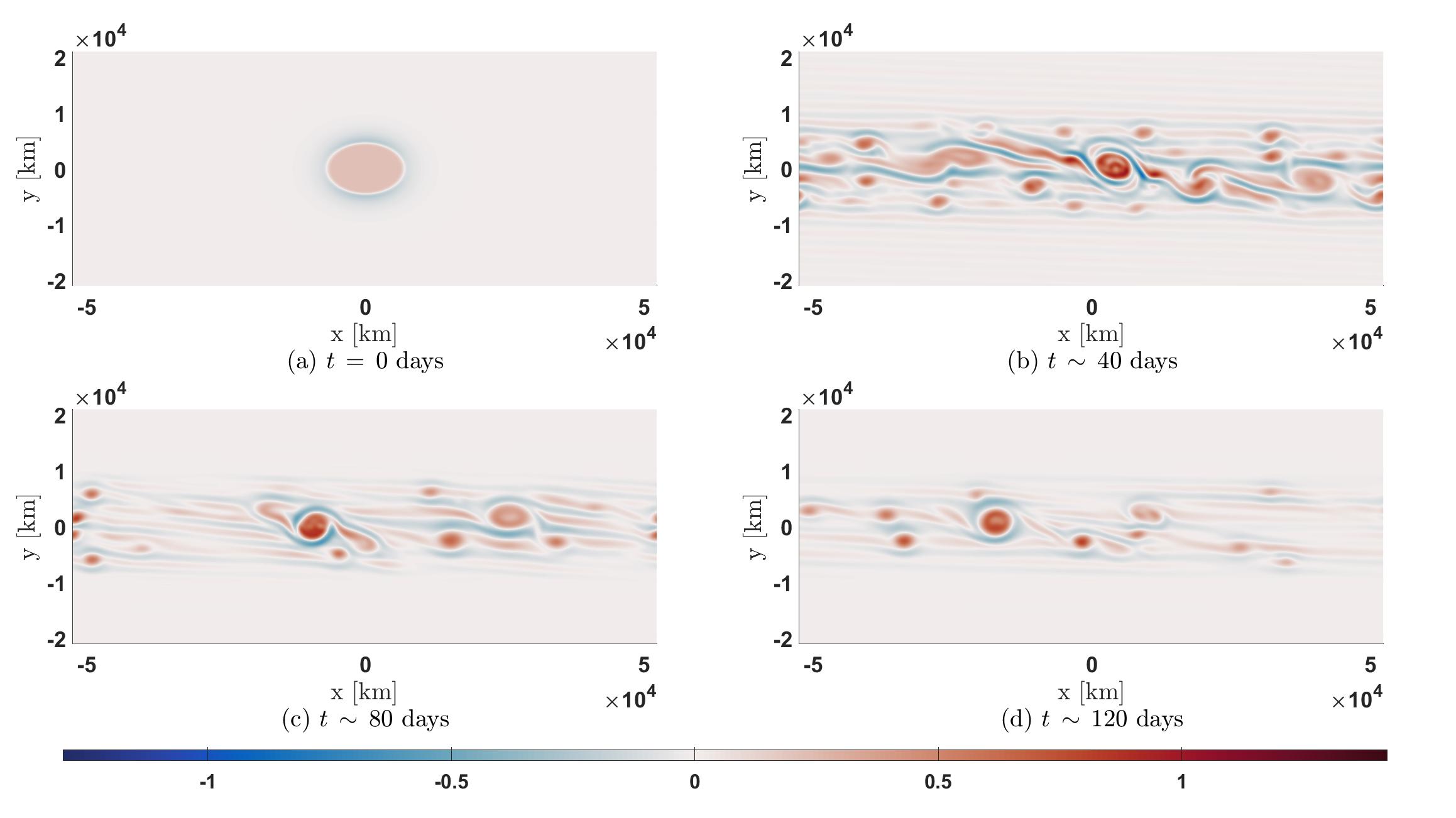}
\caption{Evolution of the $\omega_{z,vortex}(x,\,y)/|f\sin\theta_0|$ in the mid-plane of an initial $\mathrm{CA}\!-\!\mathrm{Rot}$ vortex embedded in a constant zonal shear (Case $CA1$). The surviving vortices are all part of the $\mathrm{CV}\!-\!\mathrm{Rot}$ family. (See figure \ref{fig:wz_xOz_caseCA1}.)\label{fig:wz_xOy_caseCA1}}
\end{figure}

\begin{figure}
\includegraphics[width=135mm]{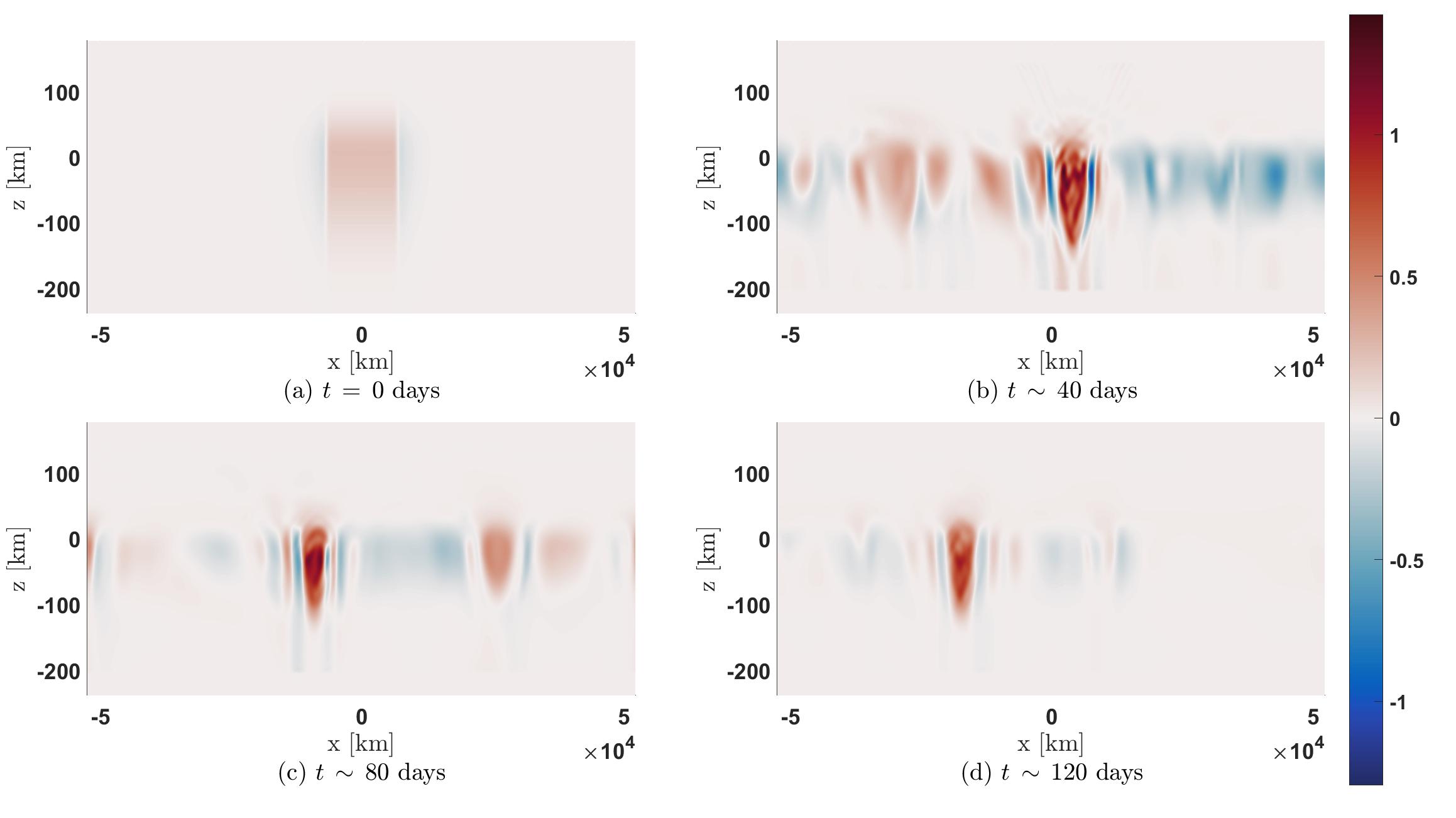}
\caption{Evolution of the $\omega_{z,vortex}(x,\,y=0,\,z)/|f\sin\theta_0|$ of the $CA1$ vortex in figure~\ref{fig:wz_xOy_caseCA1}  shown in the $x$-$z$ plane passing through the central axis of the initial vortex. The horizontal area of the initial vortex in panel (a)  is approximately constant in $z$, but the magnitude of $\omega_{z,vortex}$ decreases with distance from the mid-plane,  making it part of the $\mathrm{CA}\!-\!\mathrm{Rot}$ family. By day 40, the initial vortex and the large vortices spawned from it have changed from $\mathrm{CA}\!-\!\mathrm{Rot}$ vortices to $\mathrm{CV}\!-\!\mathrm{Rot}$ vortices. \label{fig:wz_xOz_caseCA1}}
\end{figure}

\begin{figure}
\includegraphics[width=135mm]{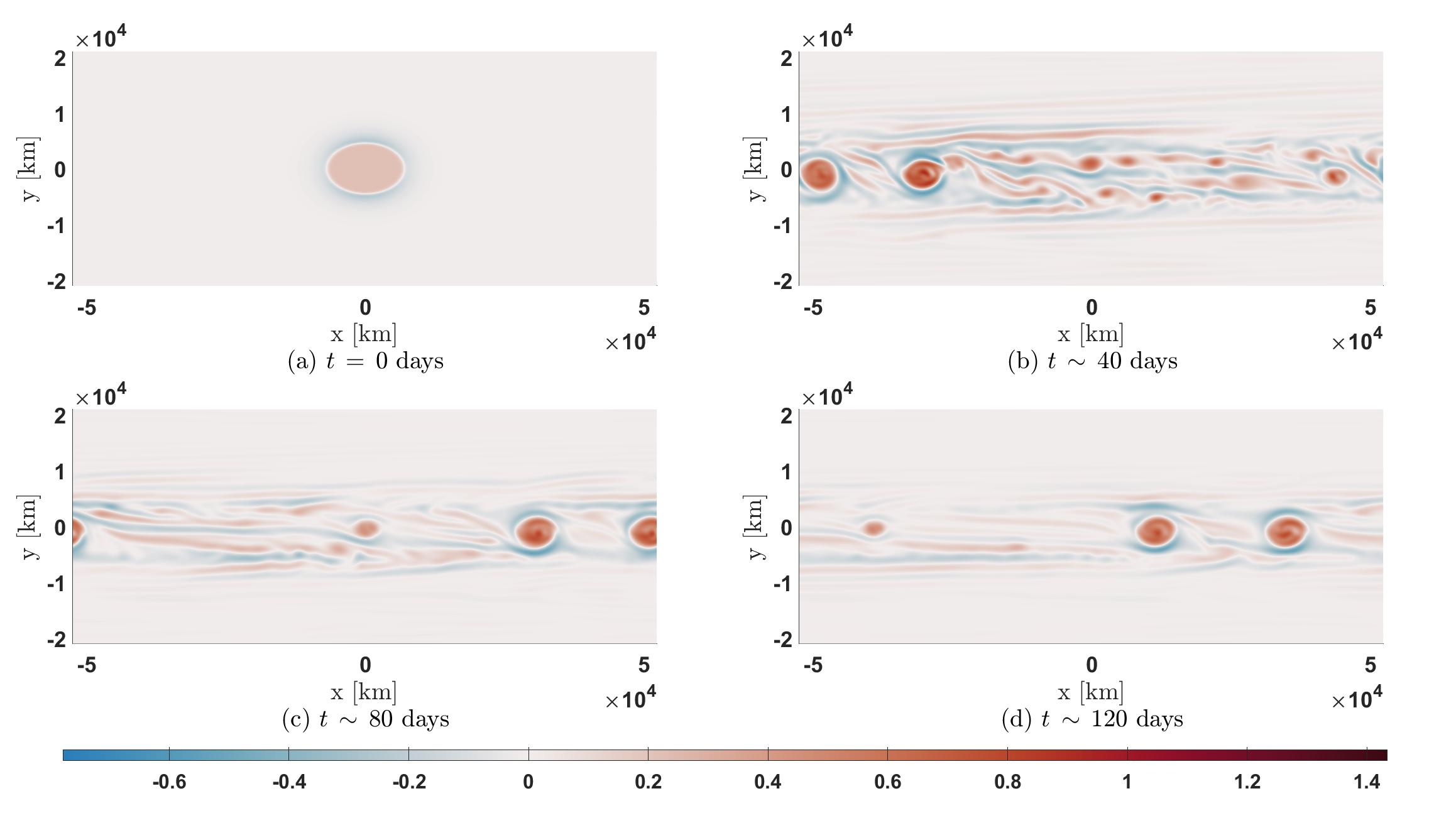}
\caption{$\omega_{z,vortex}(x,\,y)/|f\sin\theta_0|$ in the mid-plane of the same initial vortex shown in figures~\ref{fig:wz_xOy_caseCA1}, but embedded in the Jovian zonal flow. We call this the $CA2$ vortex. At 120 days, there are two large $\mathrm{CV}\!-\!\mathrm{Rot}$ vortices and one smaller one. Based on simulations of 2-dimensional quasigeostrophic vortices, we believe that the difference in the number of vortices that survive for 120 days in this figure and in figure~\ref{fig:wz_xOy_caseCA1} is not physically significant.  The initial filamentation of the out-of-equilbrium initial vortices is sensitive to small details of the initial flow, and when the filaments roll-up into vortices, their latitudes are also sensitive to the details of the initial condition. If many of the rolled-up vortices have latitudes far from the latitude of the surviving  vortex, then few of them will merge together. \label{fig:wz_xOy_caseCA2}}
\end{figure}

\begin{figure}
\includegraphics[width=135mm]{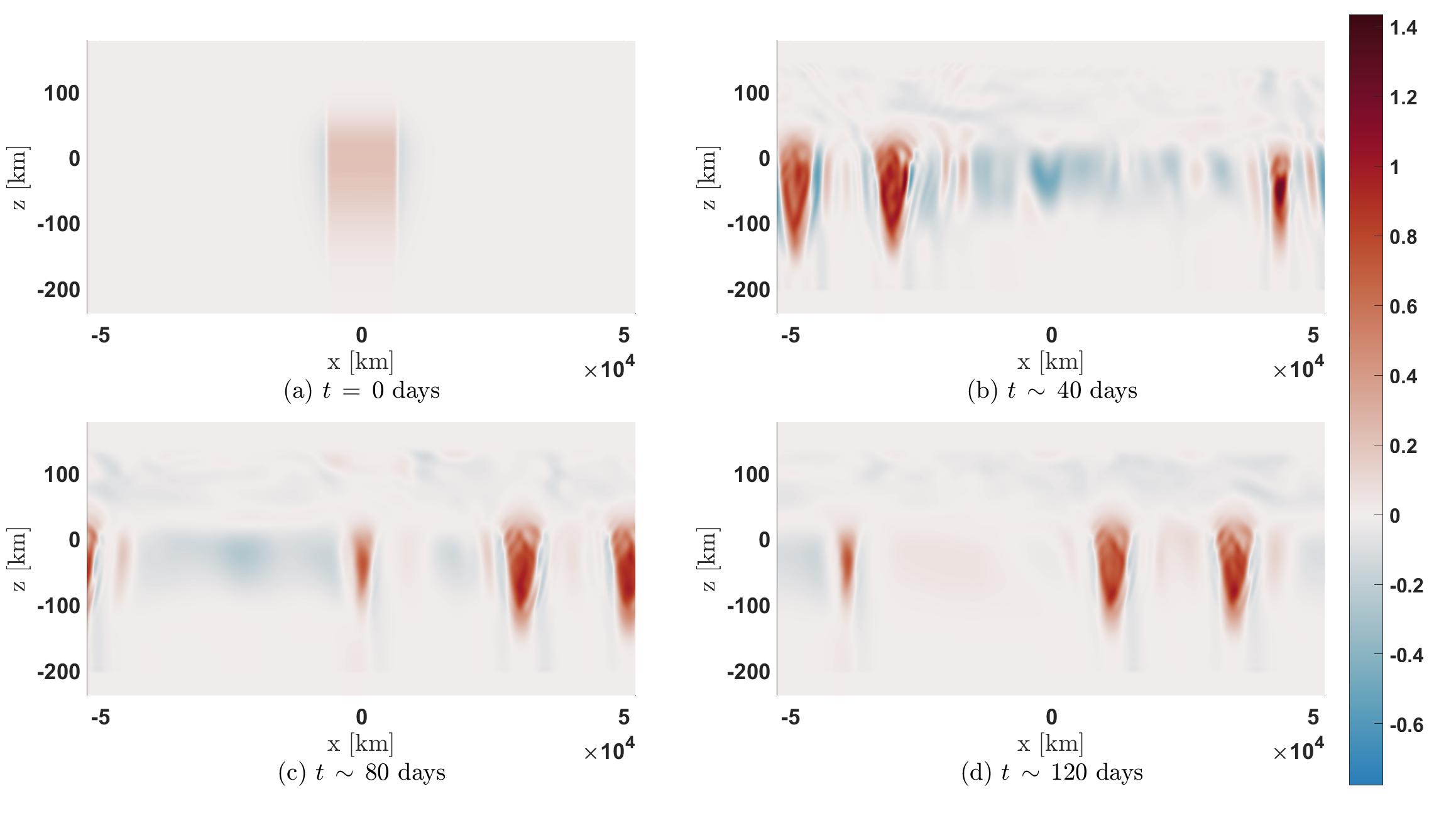}
\caption{$\omega_{z,vortex}(x,\,y=0,\,z)/|f\sin\theta_0|$ in the x-z plane passing through the central axis of the initial vortex (Case $CA2$). Although the inital vortex in panel (a) is in the $\mathrm{CA}\!-\!\mathrm{Rot}$ family, the surviving vortices in panel (d) are in the $\mathrm{CV}\!-\!\mathrm{Rot}$ family. \label{fig:wz_xOz_caseCA2}}
\end{figure}

Using equation~(\ref{eq:init_method_CA}) to create an initial $\mathrm{CA}\!-\!\mathrm{Rot}$ vortex makes the horizontal area of the vertical vorticity $\omega_{z,vortex}$ of the initial vortex the same at each height, but the magnitude of  $\omega_{z,vortex}$ decreases away from the mid-plane, vanishing at the tops and bottoms of the vortex.  
None of our simulations that began with $\mathrm{CA}\!-\!\mathrm{Rot}$ vortices produced a quasi-steady  $\mathrm{CA}\!-\!\mathrm{Rot}$ vortex at the end of the simulation. The initial vortex either quickly filamented into smaller vortices due to the shear of the zonal flow being much greater than $\omega_{z,vortex}$ at the tops and bottoms of the vortices, or the initial vortex broke apart due to a local convective instability (as discussed in $\S$~\ref{sec:numerics_convection} because the top of the initial vortex is too close to the mid-plane). In all cases, we found that the flow creates one or more quasi-steady $\mathrm{CV}\!-\!\mathrm{Rot}$ vortices at the end of the simulation.  This is shown in figures~\ref{fig:wz_xOy_caseCA1} --~\ref{fig:wz_xOz_caseCA1} for a vortex embedded in a zonal flow with constant shear (case $CA1$) and in figures~\ref{fig:wz_xOy_caseCA2} --~\ref{fig:wz_xOz_caseCA2} for a vortex embedded in the Jovian zonal flow shown in figure~\ref{fig:vzonal} (case $CA2$). For both cases, $H_{top} = 120$~km, $H_{bot} = 50$~km, $\epsilon = 1.5$ (equivalently, $Ro =-0.1$), and $z_0 = 8.2$~km (equivalently, $\bar{P}=700$~mbar). Initially, the $\mathrm{CA}\!-\!\mathrm{Rot}$ vortices are column-like in the $x$-$z$ plane with the area of the vortex on each $x$-$y$ plane identical. For both cases, the initial vortices fragment within the first 40 days into filaments that roll-up into smaller vortices. However, during that same time the initial vortex and the smaller vortices that it spawns evolve into $\mathrm{CV}\!-\!\mathrm{Rot}$ vortices. By day 80, many of the smaller vortices have merged with the large vortex, and by day 120 only a few vortices remain unmerged. We truncated the calculations after day 120 because the purpose of the calculations is to show that $\mathrm{CA}\!-\!\mathrm{Rot}$ are far from equilibrium and that they quickly evolve into $\mathrm{CV}\!-\!\mathrm{Rot}$ vortices. Based on 2-dimensional vortex dynamics in zonal flows, we expect that if we continue calculation the vortices whose central latitudes are closer together than the semi-minor radius of the largest vortex will eventually merge together.

\subsection{How the Energy Equation Comes to Equilibrium}
In $\S$~\ref{sec:IC_beginning}, we showed that our initial conditions of the vortices were such that they satisfy the continuity equation~(\ref{eq:anelastic_density}), all three components of the steady momentum equation~(\ref{eq:anelastic_momentum}), and the ideal gas equations~(\ref{eq:anelastic_ddensity}) and~(\ref{eq:anelastic_ideal_gas_law}). However because $v_z \equiv 0$, none of our initial conditions satisfied the energy equation~(\ref{eq:anelastic_energy}). Numerically, we found that with the exception of  equation~(\ref{eq:anelastic_energy}) that the full equations of motion~(\ref{eq:anelastic_density}) --~(\ref{eq:anelastic_ideal_gas_law}) were nearly satisfied 
because the initial characteristic times for ${\bfv}$, $\tilde{\rho}$, and $\tilde{P}$ to change were long compared to the turn-around time of the vortex, but the initial characteristic time for $\tilde{\Theta}$ to change was much faster.  Of the two terms on the far right side of equation~(\ref{eq:anelastic_energy}), only the $({\bfv}_{\perp} \cdot \nabla_{\perp}) \, \tilde{\Theta}$ can initially change $\partial \tilde{\Theta}/\partial t$ because $v_z$ is initially zero. When the vortex reaches its finally steady state, the magnitude of the $({\bfv}_{\perp} \cdot \nabla_{\perp}) \, \tilde{\Theta}$ term has decreased approximately by a factor of 3 from its initial magnitude and is primarily balanced by the $\bar{N}^2 v_z \, \bar{\Theta}/g$ term. Thus, even though $v_z$ is small compared to $v_{\perp}$ (see equation~\eqref{eq:vz_order_of_magnitude} below), $v_z$ cannot be zero\footnote{If $v_z=0$, then the only way that the steady version of eq.~(\ref{eq:anelastic_energy}) could be satisfied would be if the isocontours of $\tilde{\Theta}$ coincide with the 2-dimensional horizontal streamlines.} and plays an essential role in creating the final steady vortex.

\section{Scaling Analyses and the Shape of Vortices}

\label{sec:numerics_scaling}
\subsection{Scaling Analysis of the Vertical Vorticity Equation}\label{sec:numerics_scaling_wz}
In $\S$~\ref{sec:stacked_model} we showed that a steady vortex with $v_z \equiv 0$ has $(\bfv_\perp\cdot\nabla_\perp) \, (\omega_z + \beta \, y) =0$. However, none of our computed vortices have $v_z \equiv 0$, and it is important to understand how small $(\bfv_\perp\cdot\nabla_\perp) \, (\omega_z + \beta \, y)$ is, how it scales with the small dimensionless parameters of the flow, such as the Rossby number $Ro \equiv ||\omega_z||_{\infty}/(2 f  \, |\sin \, \theta_0|)$, and what terms balance it in the vertical component of the anelastic vorticity equation~\ref{eq:domegaz_dt}.
To understand how the various terms in eq.~\eqref{eq:domegaz_dt} scale, we must first understand how $\langle v_z \rangle \big / \langle v_{\perp} \rangle$ scales, where angle brackets around a quantity are defined to mean the ``characteristic value'' of that quantity. We also define the characteristic horizontal length of the vortex $L \equiv ||v_{\perp}||_2\big/||\omega_z||_2$ evaluated at the vortex mid-plane. We also define the characteristic vertical size of a vortex $H \equiv ||v_{\perp}||_2\big/||\omega_{\perp}||_2$ evaluated at the vortex mid-plane where the $L_2$ norms are taken over the horizontal area of the vortex. For the types of vortices considered here, which are not vertically confined by walls, the vertical and horizontal components of the velocity have the same vertical scale $H$, and horizontal scale $L$. In our calculations, we found that $H$ is the approximate vertical scale height of $\bar{\rho}$ and $\bar{P}$, smaller than the prescribed vertical depth $H_{top}$ or $H_{bot}$. Our result clearly shows that the vertical characteristic length scale of Jovian vortices can be different from their depth, which is similar to the findings that the Rossby deformation radius is less than the semi-minor radius of GRS \citep{marcus1988numerical,dowling1989jupiter,shetty2010changes}. 

\begin{figure}
    \centering
    \includegraphics[width=135mm]{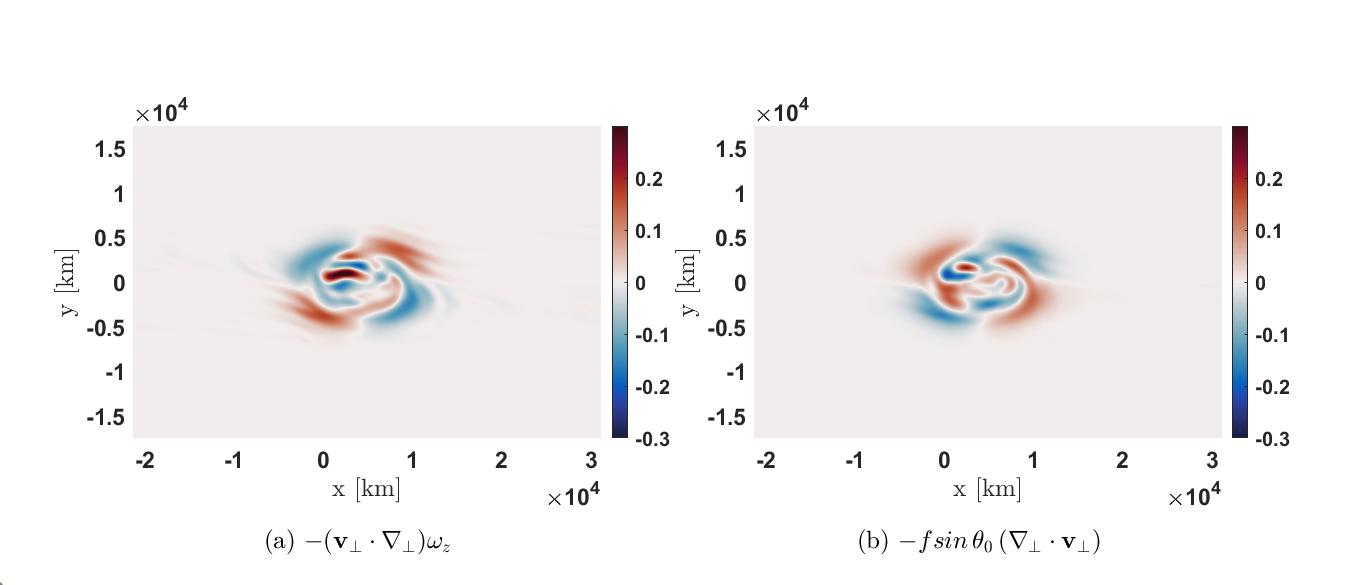}
    \caption{The dominate two components of eq.~\eqref{eq:domegaz_dt} at the vortex mid-plane (using the final state of case $CV1$). The plotted values are normalized by $\frac{||\bfv||_\infty^2}{L^2}$. Panel (a) is for  $-(\bfv_\perp\cdot\nabla_\perp) \, \omega_z$. Panel (b) is for $-f \, \sin \, \theta_0 \, (\nabla_\perp\cdot\bfv_\perp)$. }
    \label{fig:dwzdt_mid-plane_final}
\end{figure}

Note that these scaling relationships differ from those used for quasigeostrophic homogeneous flows, where it is assumed that the vertical scale height of the horizontal component of the velocity is much greater than that of the vertical component  (see Chapter 6 of Pedlosky (1982)). Our numerical calculations suggest the following scalings,

\begin{eqnarray}
\langle v_z \rangle &=& Ro^2 \, {{H}\over L}  \, \langle v_{\perp} \rangle \label{eq:vz_order_of_magnitude} \\
\Bigl\langle (\nabla_\perp\cdot\bfv_\perp) \Bigr\rangle &=&  \langle v_z \rangle/H =  Ro^2 \, \langle v_{\perp} \rangle/L \\
\Bigl\langle (\nabla \cdot\bfv) \Bigr\rangle &=&  \langle v_z \rangle/H =  Ro^2 \, \langle v_{\perp} \rangle/L  \\
\Bigl\langle (\bfv_\perp\cdot \nabla_\perp)\, (\omega_z + \beta \, y)\Bigr\rangle  &= & Ro \, \Biggl[{{\langle v_\perp \rangle}\over{L}}\Biggr]^2 \label{6}\\
\Bigl\langle f \, \sin\theta_0+ \beta \, y) \, (\nabla_\perp\cdot\bfv_\perp) \Bigr\rangle &=& Ro \, \Biggl[{{\langle v_\perp \rangle}\over{L}}\Biggr]^2\\
\Bigl\langle \omega_z \, (\nabla_\perp\cdot\bfv_\perp) \Bigr\rangle  &=& Ro^2 \, \Biggl[{{\langle v_\perp \rangle}\over{L}}\Biggr]^2 \\
\Biggl\langle v_z\pdv{\omega_z}{z}\Biggr\rangle  &=& Ro^2 \, \Biggl[{{\langle v_\perp \rangle}\over{L}}\Biggr]^2 \\
\Bigl\langle (\bfome_\perp\cdot\nabla_\perp)v_z\Bigr\rangle &=& Ro^2 \, \Biggl[{{\langle v_\perp \rangle}\over{L}}\Biggr]^2 \\
\Biggl\langle (f \, \cos_0 - \beta \, y\, \tan \theta_0) \, \pdv{v_z}{y}\Biggr\rangle  &=& Ro \,\, \cot \, \theta_0 \, {{H}\over{L}} \, \Biggl[{{\langle v_\perp \rangle}\over{L}}\Biggr]^2 \\
\Bigl\langle \beta  v_z \, \tan \, \theta_0 \Big\rangle &=& Ro \,  \, {{H}\over{r_0}} \, \Biggl[{{\langle v_\perp \rangle}\over{L}}\Biggr]^2 \label{7}
\end{eqnarray}
%
Note that the scaling equations~\eqref{6} --~\eqref{7}   
are for the seven terms on the right side of eq.~\eqref{eq:domegaz_dt}. This tells us that when the flow is steady, that although $(\bfv_\perp\cdot\nabla_\perp)\, (\omega_z + \beta y)$ is not identically zero as in eq.~\eqref{eq:b}, it is small compared to
$\Bigl[{{\langle v_\perp \rangle}\over{L}}\Bigr]^2$. Thus, the 2-dimensional streamlines and the isocontours of $(\omega_z + \beta y)$ are nearly coincident, as argued in $\S$~\ref{sec:stacked_model}. We remind the reader that this needed coincidence was the basis on which we argued that the CA family of vortices was far from equilibrium, whereas the CV family was near equilibrium. Scaling equations~\eqref{6} --~\eqref{7} also tell us that the dominant balance in eq.~\eqref{eq:domegaz_dt} is between $(\bfv_\perp\cdot\nabla_\perp)\, (\omega_z + \beta \, y)$ and $f \, \sin \, \theta_0 \, (\nabla_\perp\cdot\bfv_\perp)$.  This balance is illustrated in Fig.~\ref{fig:dwzdt_mid-plane_final}, which shows the values of $-(\bfv_\perp\cdot\nabla_\perp) \, (\omega_z + \beta \, y)$ (panel a) and 
$-f \, \sin \, \theta \, (\nabla_\perp\cdot\bfv_\perp)$ (panel b) in the mid-plane. 
These two terms do not exactly cancel each other near the center because they both become small there (due to the hollowness of the vortex), and the other terms in eq.~\eqref{eq:domegaz_dt} become comparable to them.
Note that for
most Jovian vortices and our numerical simulations here, $Ro\sim 0.1$. For the simulations presented here $H/L\sim 0.01$. The value of $H/L$ for Jovian vortices is not well constrained by observations. 
Also note that $\langle L \rangle /r_0 \simeq 0.03$ for the largest Jovian vortices.

\FloatBarrier

\subsection{Shape of the Vortex}
\begin{figure}
\begin{subfigure}{0.49\textwidth}
\includegraphics[width=66mm]{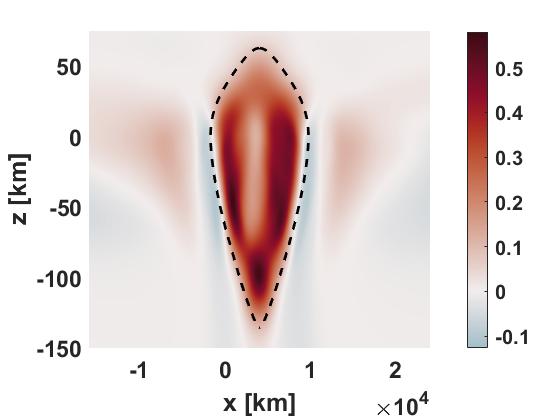}
    \caption{Case $CV1$}
    \label{fig:wz_fit_CV1}
\end{subfigure}
\begin{subfigure}{0.49\textwidth}
\includegraphics[width=66mm]{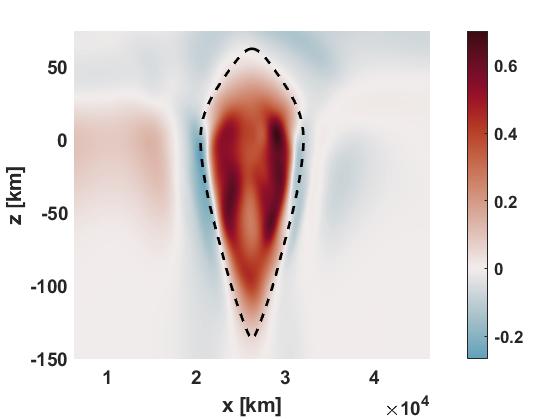}
    \caption{Case $CV2$}
    \label{fig:wz_fit_CV2}
\end{subfigure}
\caption{Comparing $2\sqrt{{\cal A}(z)}$ (dashed line) to the vertical vorticity $\omega_{z,vortex}(x,\,y=0,\,z)/|f\sin\theta_0|$ of the  $CV1$ and $CV2$ vortices in the $x$-$z$ plane going through the center of each vortex. The factor of 2 is to make the boundary close to the visual boundary of the vortex.\label{fig:wz_fit}}
\end{figure}

The shapes of the $\omega_{z,vortex}$ of the steady vortices in the $x$-$z$ plane all have a distinctive ``ice cream cone'' shape (e.g.,  panel (d) in Figs.~\ref{fig:wz_xOz_caseCV1}, \ref{fig:wz_xOz_caseCV2}, \ref{fig:wz_xOz_caseCA1}, and \ref{fig:wz_xOz_caseCA2}). We can explain these by recalling that 
\begin{eqnarray}
N^2\equiv\frac{g}{\bar{\Theta}}\pdv{\Theta}{z}=\bar{N}^2+\frac{g}{\bar{\Theta}}\pdv{\tilde{\Theta}}{z}, \label{eq:full_N2}
\end{eqnarray}
and using the hydrostatic approximation~\eqref{eq:near_hydrostatic_eq},
\begin{eqnarray}
N^2&=&\bar{N}^2+\frac{g}{\bar{\Theta}}\pdv{}{z}\,\Bigg[\frac{\bar{\Theta}}{g}\pdv{}{z}\,\Big(\frac{\tilde{P}}{\bar{\rho}}\Big)\Bigg]\\
&=&\bar{N}^2+\dv{ln\bar{\Theta}}{z}\pdv{}{z}\,\Big(\frac{\tilde{P}}{\bar{\rho}}\Big)+\pdv[2]{}{z}\,\Big(\frac{\tilde{P}}{\bar{\rho}}\Big) \label{eq:N2_full_comp}
\end{eqnarray}
To eliminate $\Bigl(\frac{\tilde{P}}{\bar{\rho}}\Bigr)$ in eq,~\eqref{eq:N2_full_comp}, we use the small-Rossby number limit (quaisgeosgtrophic approximation) of equation \eqref{eq:c} with $\beta=0$
\begin{eqnarray}
(f \, \sin\, \theta_0) \, \omega_z(x, y, z) = \nabla_\perp^2 \, \Biggl(\frac{\tilde{P}(x, y, z)}{\bar{\rho}}\Biggr). \label{new}
\end{eqnarray}
For each $z$ within the vortex, averaging eq.~\eqref{new} over the horizontal area within the vortex we define ${\cal A}(z)$ with
\begin{eqnarray}
(f \, \sin\, \theta_0) \, \omega_z(z)\Big|_{averaged} =  \Biggl[\nabla_\perp^2 \, \Biggl(\frac{\tilde{P}}{\bar{\rho}}\Biggr)\Biggr]\Bigg|_{averaged} \equiv - \Biggl[\Biggl(\frac{\tilde{P}}{\bar{\rho}}\Biggr)\Bigg|_{averaged}\Biggr] \Bigg/ {\cal A}(z), \label{newer} 
\end{eqnarray}
where ${\cal A}(z)$ has the dimension of a length squared. Our numerical calculations show that $\sqrt{{\cal A}(z)}$ is approximately equal to the mean cross-sectional radius of the vortex.\footnote{We first empirically found this relationship in Boussinesq vortices \citep{hassanzadeh2012universal}.}  Because all the final, statistically-steady vortices computed here, regardless of their initial conditions, are $\mathrm{CV}\!-\!\mathrm{Rot}$ vortices, it is reasonable to approximate $\omega_z(z)|_{averaged}$ as independent of $z$ and assign it a value $\omega_A$. Thus, eqs.~\eqref{eq:N2_full_comp} --~\eqref{newer} become
\begin{eqnarray}
\dv[2]{{\cal A}}{z}+\dv{ln\bar{\Theta}}{z}\dv{{\cal A}}{z}=-\frac{N^2(z)\big|_{averaged}-\bar{N}^2(z)}{f \, \omega_A \, \sin \, \theta_0 } \label{eq:N2_to_A}
\end{eqnarray}
Given
$dln\bar{\Theta}/dz$, $\bar{N}^2(z)$, $N(z)|_{averaged}$, and $\omega_A$, we can solve equation~\eqref{eq:N2_to_A} for $\sqrt{{\cal A}(z)}$, the mean radius of the vortex, using the boundary conditions that ${\cal A} =0$ at the top and bottom of the vortex. Figure~\ref{fig:wz_fit} shows the results of solving equation~(\ref{eq:N2_to_A}) for $\sqrt{{\cal A}(z)}$ for the $CV1$ and $CV2$ vortices, and that the equation works well in describing the vortex shapes.

\par

\section{Constraints Due to Local Convective Instabilities}\label{sec:numerics_convection}
We have found that if an initial vortex has any location in which $N(x, y, z)$ is imaginary, or equivalently, $\partial \Theta(x, y, z) /\partial z \le 0$ that the flow is, as expected, locally unstable to convection. We can show that the convective instability limits how close the tops and bottoms of vortices can be to the vortex mid-planes.
\begin{figure}
\begin{subfigure}{0.49\textwidth}
\includegraphics[width=62mm]{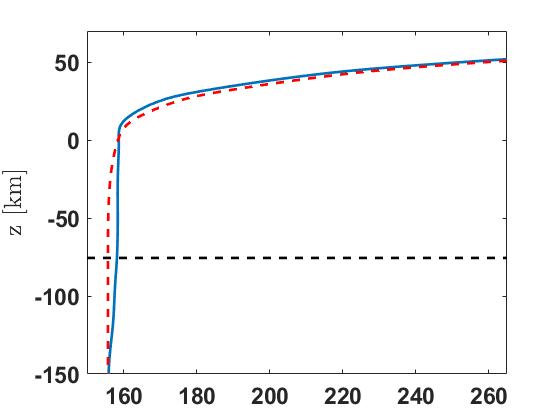}
    \caption{potential temperature $\Theta\,[K]$}
    \label{fig:ptempzfull}
\end{subfigure}
\begin{subfigure}{0.49\textwidth}
\includegraphics[width=62mm]{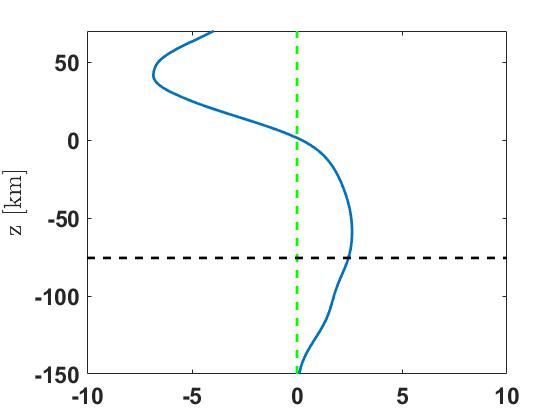}
    \caption{potential temperature anomaly $\tilde{\Theta}\,[K]$}
    \label{fig:ptemp_z_tilde}
\end{subfigure}
\caption{The potential temperature of an anticyclone ($CV1$). The horizontal black dashed line is the top of the convection zone at $z_{convection} =10$~bar. Panel a: The blue curve is $\Theta(z)$ along the ${\bf z}$ the center axis of the vortex. The dotted red curve is $\bar{\Theta}(z)$. $z_{cross}$ is the height near the mid-plane where the two curves cross. Panel b: The blue curve is $\tilde{\Theta}(z)$, and $z_{cross}$ is the height near the mid-plane where $\tilde{\Theta}=0$ on the blue curve. The green dashed line is where $\tilde{\Theta}=0$.}
\end{figure}

\begin{figure} 
\includegraphics[width=135mm]{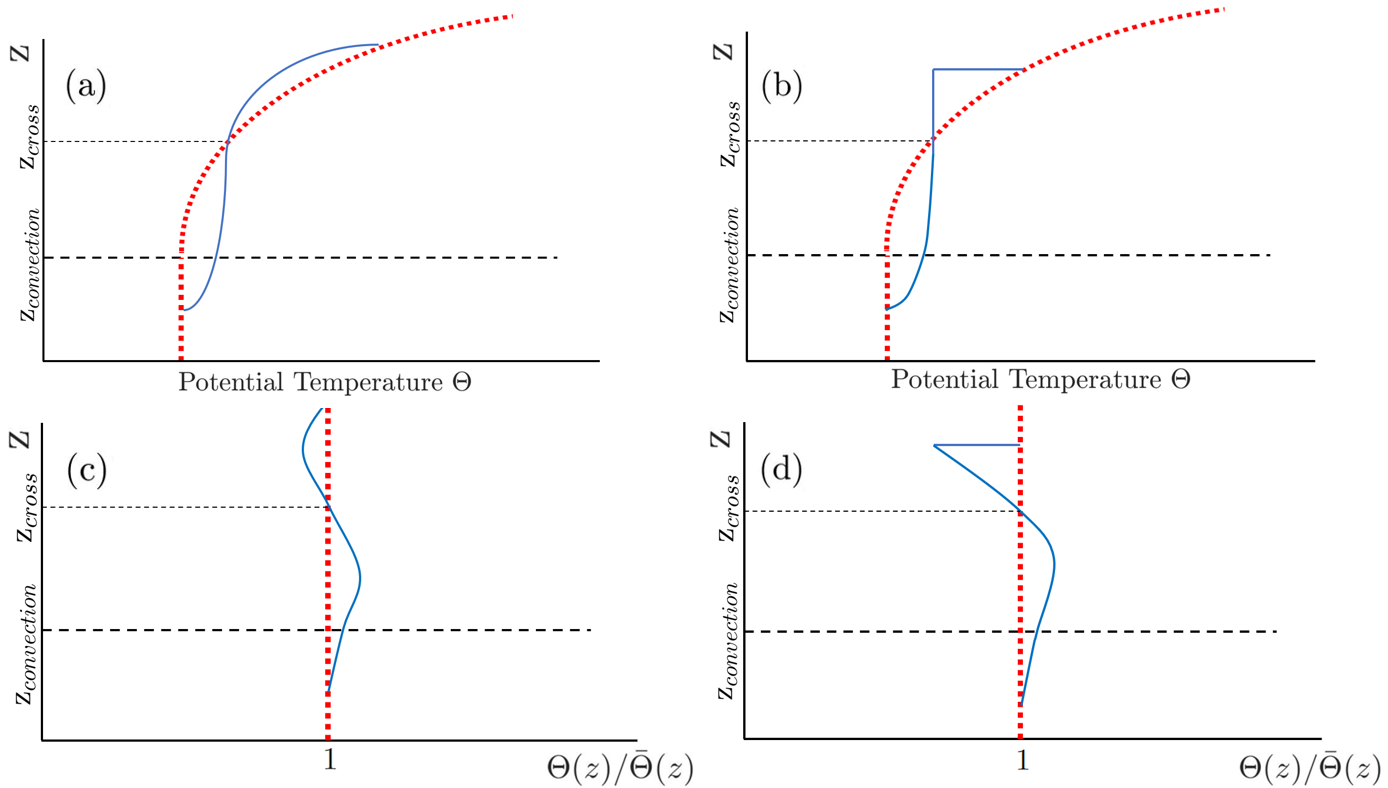}
\caption{Schematic showing the minimum possible height of the top of the vortex $z_{top}$ without being convectively unstable. See text for details. The long-dashed horizontal lines show the top of the convective zone at $z_{convection}$; the short-dashed horizontal lines show $z_{cross}$.  The values of $z_{cross}$ and $z_{convection}$ are the same in all four panels.  Panels a and b: The red dotted curves are $\bar{\Theta}(z)$; the blue solid curves are $\Theta(x, y, z)$ at some fixed $(x, y)$ within the vortex.  Panel a is  for a vortex with $z_{top}$ greater than the minimum value; panel b is for the vortex with the same flow at $z < z_{cross}$ as in (a), but with a minimum $z_{top}$.  Local convective stability requires that the slope $\partial \Theta/\partial z  \ge 0$. Panels c and d: The vertical red dotted curves are $\Theta(z)/\bar{\Theta}(z) \equiv 1$; the blue solid curves are $\Theta(z)/\bar{\Theta}(z)$ at $(x, y)$. Panel c is for the vortex in (a), and panel d is for the vortex in (b). In (c) and (d), hydrostatic equilibrium requires that the area of the lobes at $z > z_c$ equal the areas of the lobes at $z< z_c$.
  \label{fig:schematicpv}}
\end{figure}

For all $x$ and $y$ locations within a vortex, we define $z_{top}(x, y)$ and $z_{bottom}(x, y)$ to be the top and bottom of the vortex, where the tops and bottom are the locations where 
$\tilde{P}(x, y, z)$ is approximately zero. We define 
$z_{cross}(x, y)$ to be the location near the mid-plane where $\tilde{\Theta}(x, y, z) = 0$ 
(where the red and blue lines cross in 
Fig.~\ref{fig:ptempzfull}.
The requirement that 
\begin{eqnarray}
N^2(x,y,z)=\frac{g}{\bar{\Theta}}\pdv{\Theta}{z}\geq0 \label{convection}
\end{eqnarray}
at all locations in the fluid along with the hydrostatic pressure equation put bounds on $z_{top}(x, y)$ and $z_{bottom}(x, y)$. Two integral relationships of hydrostatic equilibrium can be determined by integrating the approximate anelastic hydrostatic equation~\eqref{eq:near_hydrostatic_eq} from the bottom of the vortex at $z_{bottom}(x, y)$ to $z_{cross}(x, y)$ and from $z_{cross}(x, y)$ to the top of the vortex at $z_{top}(x, y)$.
\begin{eqnarray}
\tilde{P}(x, y, z_{cross})\big/[g \bar{\rho}(z_{cross})] &=&  \int_{z_{bottom}}^{z_{cross}} \, dz \,\, \tilde{\Theta}(x, y, z_)\big/\bar{\Theta}(z)  \label{hbottom} \\
\tilde{P}(x, y, z_{cross})\big/[g \bar{\rho}(z_{cross})] &=&  -\int_{z_{cross}}^{z_{top}} \, dz \,\, \tilde{\Theta}(x, y, z)\big/\bar{\Theta}(z).  \label{htop}
\end{eqnarray}
Figure~\ref{fig:schematicpv}a schematically shows the potential temperature $\Theta$ of an anticyclone similar to the one in Figure~\ref{fig:ptempzfull}. Figure~\ref{fig:schematicpv}c shows the corresponding $\Theta/\bar{\Theta}$. Equations~\eqref{hbottom} and~\eqref{htop} show that the areas of the two lobes (between the vertical dotted red line and the solid blue curve in Figure~\ref{fig:schematicpv}c) above $z_{cross}$ and below $z_{cross}$ are the same. Let us modify the vortex in \ref{fig:schematicpv}a for $z > z_{cross}$ in order to minimize $z_{top}$, while keeping the vortex at $z < z_{cross}$ unchanged, and therefore the area of the lower lobe in figure \ref{fig:schematicpv}c unchanged. The modified vortex is shown in figures~\ref{fig:schematicpv}b and~\ref{fig:schematicpv}d. To maintain hydrostatic equilibrium, the areas of the upper and lower lobes in the modified vortex in Figure~\ref{fig:schematicpv}d must also be the same. 
The modified vortex in figure~\ref{fig:schematicpv}b must have $\partial \Theta/\partial z \ge 0$ to be stable, but to minimize $z_{top}$, it needs to have  $\partial \Theta/\partial z$ as small as possible. 
Thus, the vortex with the minimum $z_{top}$ has $\partial \Theta/\partial z  =0$ at $z \ge z_{cross}$.  Our numerical simulations with initial  vortices with a $z_{top}$  smaller than the minimum value implied by figure~\ref{fig:schematicpv} show that the vortices  either break apart or increase the heights of their tops as they evolve in time. A similar argument to the one illustrated in Fig.~\ref{fig:schematicpv} shows that there is maximum upper bound to $z_{bottom}$, and we have validated that argument numerically.

\begin{figure} 
\includegraphics[width=135mm]{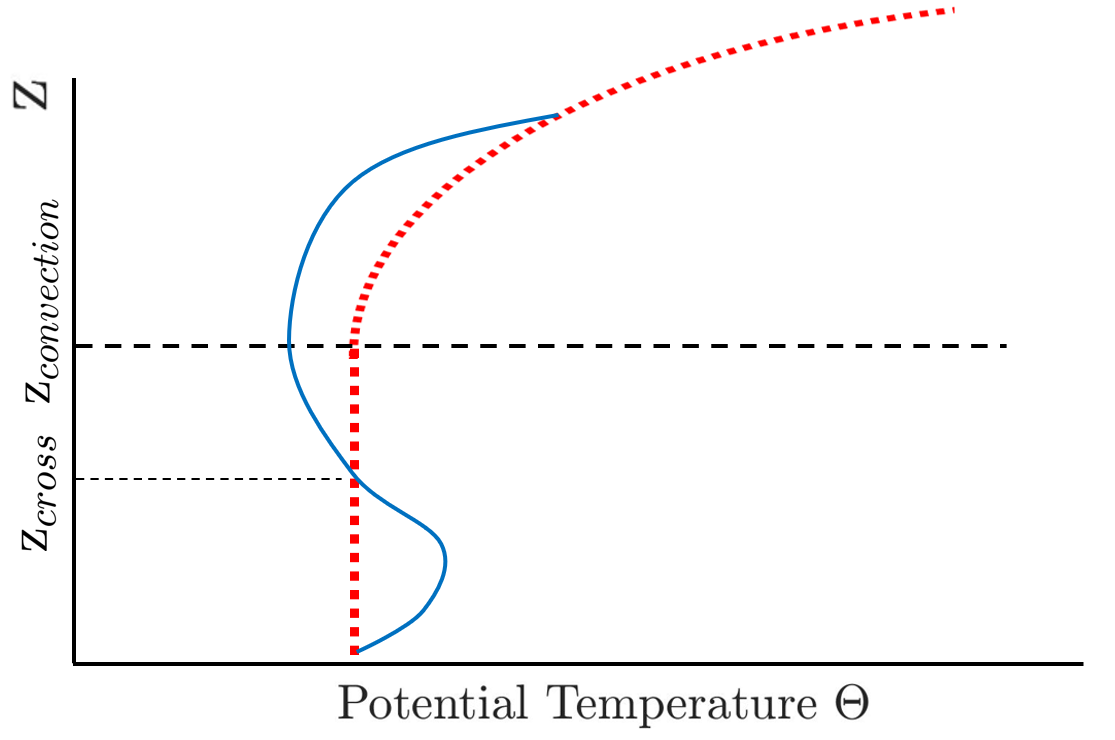}
\caption{Schematic with curves for $\Theta(z)$ and $\bar{\Theta}(z)$, and lines for $z_{cross}$,  and $z_{convection}$ plotted as in Fig.~\ref{fig:schematicpv}a. If $z_{cross} < z_{convection}$, then $\Theta(z_{convection}) < \Theta(z_{cross})$, so $\partial \Theta /\partial z < 0$ somewhere in the range $z_{cross} < z < z_{convection}$, and the vortex is locally unstable to convection.}
  \label{fig:impossible}
\end{figure}

Figure~\ref{fig:impossible} is a schematic like Fig.~\ref{fig:schematicpv}a. It shows that $z_{cross}$ must be greater than $z_{convection}$ for all $x$ and $y$. If $z_{cross} < z_{convection}$ as pictured in figure~\ref{fig:impossible}, then at  $z_{convection}$, $\Theta < \bar{\Theta}$.  This means that $\partial \Theta /\partial z <0$ for some values of $z$ in the range $z_{cross} < z < z_{convection}$, making the flow locally unstable to convection. Thus, $z_{cross} > z_{convection}$. Because our simulations show that $z_{cross}$ is approximately the height of the mid-plane of the vortex (see figure~\ref{fig:ptemp_xOz_caseCV1}), this suggests that the mid-plane is always at a height above the convection zone.

\section{Conclusions, Discussion and Future Work}

\subsection{Summary}
Although the main objective of this study was not to replicate any specific planetary vortex, our simulations were carried out in the anticipation that the stable, 3-dimensional vortices would qualitatively resemble Jovian vortices, including the Great Red Spot (GRS). Our focus is on understanding the vertical structure of planetary vortices because their horizontal structures are often well-constrained by observations (but typically at only one height in the atmosphere), whereas their vertical structures remain unobserved and poorly understood.  Our main results are:
\begin{itemize}
\item \hspace{0.2in} This is the first numerical calculation of Jovian anticyclonic vortices with temperatures and horizontal velocities similar to the GRS that are solutions to the 3-dimensional, anelastic equations of motion. The vortices are computed with the observed atmospheric vertical thermal stratification or Brunt-V\"{a}is\"{a}l\"{a} frequency $\bar{N}(z)$ so that the  atmosphere is strongly stratified at its top and nearly adiabatic at is bottom. The calculations use the observed (cloud-top level) east-west zonal shear flow. The vortices are shielded (i.e., no net circulation of the vertical vorticity) and their heights of their tops and bottoms are within observational bounds. The bottoms of the stable vortices that we computed were at approximately 30~bar or 150~km beneath the top of the visible clouds. 
The horizontal velocity field at the cloud-top level (the only height at which we have direct observations) qualitatively agree with observations. The calculations were computed with no dissipation other than hyperviscosity and hyperdiffusivity (i.e., no viscosity, radiative transfer terms, thermal conductivity, or Newton cooling) and after 120 days, the vortices do not appear to decay with time. 
\item \hspace{0.2in} One of the reasons we are able to compute these vortices is that the initial-value code used to integrate the governing equations can take large time steps with no loss of accuracy or stability due to its novel-time stepping scheme that is an improvement upon the Krylov-like method of \citet{barranco20063d}. (See Appendix.)
\item \hspace{0.2in} There is a stable family (that we call $\mathrm{CV}\!-\!\mathrm{Rot}$) of vortices in which the vertical vorticity is nearly uniform  in the direction parallel to the planet's spin axis, but the horizontal cross-sectional area of the vortex varies with the distance in that direction. The top and bottom of the vortices are the locations where the cross-sectional area goes to zero, rather than  locations where the vertical vorticity goes to zero. 
\item \hspace{0.2in} An alternative proposed family ($\mathrm{CA}\!-\!\mathrm{Rot}$) of vortices \citep{galanti2019determining,parisi2021depth} has its horizontal cross-sectional area nearly constant in the direction parallel to the planet's spin axis, but the magnitude of the vertical vorticity varies in that direction. The tops and bottoms of the vortices are the locations where the vertical vorticity goes to zero, rather than the locations where horizontal cross-sectional area goes to zero. We have analytically shown that
these proposed vortices are far from equilibrium.  The initial-value code shows that the $\mathrm{CA}\!-\!\mathrm{Rot}$ vortices  break apart and re-assemble as one or more $\mathrm{CV}\!-\!\mathrm{Rot}$ vortices.
\item \hspace{0.2in} We develop an algorithm for creating families of vortices that are close to equilibrium.  Many properties of the vortices, such as the heights of their tops, bottoms, etc., can be specified, as well as their horizontal velocities at the cloud-top levels. When the initial-value code is initialized with a vortex from the  $\mathrm{CV}\!-\!\mathrm{Rot}$ family, it often comes to a final quasi-steady state close to its initial conditions.  Thus, there is an easy way to produce long-lived, Jovian-like vortices.
\item \hspace{0.2in} The computed anticyclonic vortices have cool-tops (approximately $5K$), which is similar to the $\sim8K$ cool-top observations \citep{hanel1979infrared,flasar1981thermal,orton1996galileo,sada1996comparison,simon2002new,fletcher2010thermal,orton2022juno}) and warm-bottoms with respect to the surrounding atmosphere, similar to Jovian anticyclones. 
\item \hspace{0.2in} An unexpected result is that initially non-hollow vortices evolve to quasi-steady hollow vortices, like the GRS (i.e., with a high-speed thin annulus at the outer edge of the vortex and a quiet center with a minimum of vorticity).
\item \hspace{0.2in} A second unexpected result is that at the outer edges of quasi-steady final anticyclonic vortices, there is a  thin annular region that is warmer than the ambient fluid. The initial conditions do not have this annular temperature anomaly; it is created when a down-welling flow (i.e., $v_z < 0$) forms spontaneously within the annulus. These warm outer annuli might explain the rings of bright infrared emissions that surround the GRS and many of the other Jovian anticyclones observed by \citet{de2010persistent} and \citet{fletcher2010thermal}.
\item \hspace{0.2in} We find new scaling laws for vortices in a vertically unbounded rotating, stratified atmosphere (rather than that confined vertically in a thin shell as in Pedlosky (1987), Chapter 6, i.e, so that the vertical scale height of the horizontal flow is much larger than that of the vertical flow). We show that the vertical scale $H$  of the horizontal and vertical components of the velocity are the same. We determine how the ratio of the magnitudes of the vertical-to-horizontal velocities scale with Rossby number $Ro$ and $H/L$, where $L$ is the horizontal scale of the velocities. We determine how the divergence of the velocity and other properties of the flow scale with $Ro$ and $H/L$.
\item \hspace{0.2in} Vortices whose central axes are initially aligned with the local gravity (i.e., aligned with the spherical radial direction of the planet),
evolve so that their central axes are approximately aligned with the planet's axis of rotation.
\item \hspace{0.2in} Although we do not compute whether the bottom of a vortex can survive when it is within a fully turbulent convection zone, we prove that the mid-plane of the vortex must lie at a height above the top of the convective zone. 
\item \hspace{0.2in} The cross-section of the vortex in the vertical-horizontal plane has an ``ice cream cone'' shape that can be explained by a simple expression.
\item \hspace{0.2in} The magnitudes of the Rossby numbers in parts of the Jovian-like vortices that we have simulated are approximately from $-0.3$ to $-0.1$. so if quasigeostrophic approximations are used to model these vortices, the approximations would only be expected to be accurate to within 30\%. In particular, a comparison of the thermal wind equation with the curl of the momentum equation, from which it is derived, shows that if the thermal wind equation is used to approximate the relationship between the horizontal temperature gradients within the vortices and the vertical gradients of the horizontal velocities, those approximations are only accurate to within 30\%.
\end{itemize}

\subsection{Discussion and Future Work}

We recognize that our study is not exhaustive and that there may be other stable families of Jovian-like vortices. Our calculations do not include $\beta$ effects or a fully turbulent convective zone, and we shall address those problems in future calculations because we do not anticipate that they will be computationally challenging. Micro-scale physics, such as the formation and sublimation of clouds with its associated diabatic heating, etc., are not included in this study, which might be important. We plan to examine some of these effects by incorporating a subgrid-scale clould model in our future calculations. 

The calculations presented here use a zonal east-west flow that does not vary in the vertical direction. It is set to the observed zonal flow at $\sim$~1~bar. In future calculations we propose to make it a function of height, but currently there are no consistent observational data to guide our choices \citep{fletcher2021jupiter}. However, we believe that future observations, using the James Webb Space Telescope \citep{de2022jwst}, will provide  zonal velocities at another height,  between 50 and 700~mbar in the atmosphere, to guide us and also to provide  horizontal velocity fields of the GRS and other Jovian vortices which can be used to test our numerical calculations.

The focus of our future calculations will be to determine under what circumstances and for what parameter values do initial vortices in the $\mathrm{CV}\!-\!\mathrm{Rot}$ family evolve into long-lived,  quasi-steady vortices. For those vortices that do become quasi-steady, what properties (e.g., thickness of the annular shield, location of the top, horizontal area, height of the mid-plane) of the initial vortex are inherited by the final quasi-steady vortex, which change, and why? Can we initialize the flow with a large-area $\mathrm{CV}\!-\!\mathrm{Rot}$ vortex that evolves into a long-lived vortex as large as the GRS and with a horizontal velocity field at $\sim$~1~bar that quantitatively matches it? Can we start with small-area $\mathrm{CV}\!-\!\mathrm{Rot}$ vortices that evolve into long-lived vortices with sizes and horizontal velocity fields at $\sim$~1~bar that quantitatively match the current Red Oval or any of the three previous White Ovals, which are not hollow? We also plan to add radiative transfer terms to the energy equation to mimic the $\sim$~4.5-year thermal dissipation time of the atmosphere near 1~bar to determine whether the numerically-computed Jovian-vortices are still long-lived.

\section*{Acknowledgements}
We acknowledge the use of NSF ACCESS computational resources (ID: PHY220056). We acknowledge the use of observation data and helpful comments from Imke de Pater, Chris Moeckel, Michael Wong, Haley Wohlever, Anton Ermakov, and Amy Simon.\par

\section*{Declaration of Interests.} 
The authors report no conflict of interest.

\section*{Appendix: Numerical Method}
\label{sec:numerical_method}
Equations \eqref{eq:anelastic_density} to \eqref{eq:anelastic_ideal_gas_law} are numerically solved with a 3-dimensional pseudo-spectral code. The code is based on one that we developed for  Boussinesq studies of ocean vortices \citep{hassanzadeh2012universal,mahdinia2017stability} and anelastic studies of protoplanetary disks \citep{marcus2013three, marcus2015zombie,marcus2016zombie,barranco2018zombie}. The time-stepping algorithm uses different fractional steps for the nonlinear and linear terms in the equations of motion. The fractional steps for the nonlinear terms use an Adams-Bashforth method, but the fractional steps for the linear terms is not based on a Taylor series expansion, but rather, a direct form of exponentiation like a Krylov method. The time-stepping algorithm differs from the one used by \citet{barranco20063d}, by suppressing a spurious solution that grows when the time step is too large. In regions where the flow is strongly stratified and $N/f \gg 1$, the criterion for a stable (and accurate) time step $\Delta t$ is not the Courant condition, but rather the requirement that $N \, \Delta t$ is sufficiently small throughout the computational domain. In our simulations, the maximum value of $\bar{N}$ is $\bar{N}_{max} =  2.4 \times 10^{-2}\mathrm{s}^{-1}\simeq\,180 f \, \sin \, \theta_0$. The time step $\Delta t$ we use makes  $\bar{N}_{max} \,  \Delta t \simeq 2.2$, which is a much larger value than used in most other initial-value numerical calculations of stratified flows.  However, we note the Coriolis term $(f\sin\theta_0) \, \Delta t \simeq0.012$ is much smaller than this value. One test that we used to determine the accuracy of our initial-value code was to compute the dispersion relation for Poincar\'e waves in an isothermal atmosphere. We compared the dispersion relationship that we obtained using our initial-value code to the relationship that we obtained semi-analytically by computing the eigenmodes of the linearized equations of motion. Using the same $\Delta t$ in the initial-value code that we used to compute the Jovian vortices, we found that the fractional difference between the two dispersion relations was less than $10^{-3}$.

All simulations use 384 Fourier modes in the $x$ and the $y$ directions, and 385 Chebyshev modes in the vertical  $z$ direction. The numerical code has a spatial resolution of $\Delta x\,=\,270$~km, $\Delta y\,=\,180$~km, $max(\Delta z)\,=\,1.7$~km. The domain is periodic in the $x$ direction, and  $v_z =0$ at the vertical boundaries. The computational box size in the $x$ direction is as shown in figure~\ref{fig:wz_xOy_caseCV1} with $|x| \le 5.25 \times 10^4$~km. Figure~\ref{fig:wz_xOy_caseCV1} also displays the computational results for $|y| \le 2 \times 10^4$~km, which is the region of physical interest to us in examining the vortex dynamics. However, throughout this paper we use a computational domain in $y$ that is larger, and is  $|y| \le 3.5 \times 10^4$~km. 
Similarly, the computational domain in the vertical direction is larger than the domain of physical interest.
We use larger computational domains for two reasons. The figures display the results that we believe to be physically relevant. The undisplayed results lie in regions in which we have included sponge layers where the anomalous velocity $\bfv_{vortex}$ and the anomalous potential temperature $\tilde{\Theta}_{vortex}$ are damped to zero via Rayleigh drag and Newton cooling, respectively \citep{mahdinia2017stability} to dissipate any internal gravity waves that enter them. Without the sponge layers, the internal gravity waves would be reflected and contaminate the calculation. The sponge layers at the $y$ and $z$ boundaries are $7,000$~km and $31$~km thick, respectively.

The second reason that the computational domain in $y$ that is larger than the physically-relevant domain shown in figure~\ref{fig:wz_xOy_caseCV1} and elsewhere is that the calculations within the computational $y$ domain are periodic in $y$, but the zonal flow $v_{zonal}(y)$ is not periodic. Therefore in the regions outside the physically relevant domain, we artificially modify $v_{zonal}(y)$ to make it periodic in $y$. This is a well-known and very versatile computational trick, which does not seem to affect the results. (See \citet{marcus1993jupiter}.) 

\newpage
\bibliographystyle{jfm}
\bibliography{bib_jupiter.bib}

\begin{thebibliography}{90}
\bibitem[Cassini(1666)]{cassini1more}
{\sc \au{Cassini, Jean~Dominique}} \yr{1666}  \at{A more particular account of
  those observations about jupiter, that were mentioned in numb. 8}.
  \jt{Philosophical Transactions of the Royal Society of London}
  \bvol{1}~(10),  \pg{171--173}.
\end{thebibliography}

\end{document}